\documentclass[journal,draftclsnofoot,12pt,onecolumn]{IEEEtran}
\usepackage{graphicx}
\usepackage{amssymb}
\usepackage{subfigure}
\usepackage{caption}
\renewcommand{\theequation}{\arabic{section}.\arabic{equation}}
\newtheorem{theorem}{Theorem}

\newtheorem{lemma}{Lemma}

\newtheorem{remark}{Remark}

\begin{document}

\title{Feedback Coding Schemes for the Broadcast Channel with Mutual Secrecy Requirement}

\author{Bin~Dai,
        Linman~Yu,
        and~Zheng~Ma
\thanks{B. Dai is with the
School of Information Science and Technology,
Southwest JiaoTong University, Chengdu 610031, China, and with
the State Key Laboratory of Integrated Services Networks, Xidian University, Xi$'$an, Shaanxi 710071, China,
e-mail: daibin@home.swjtu.edu.cn.}
\thanks{L. Yu is with the School of Economics and Management, Chengdu Textile College, Chengdu 611731, China,
Email: yulinmanylm@163.com.}
\thanks{Z. Ma is with the
School of Information Science and Technology,
Southwest JiaoTong University, Chengdu, China,
e-mail: zma@home.swjtu.edu.cn.}
}

\maketitle

\begin{abstract}

Recently, the physical layer security (PLS) of the communication systems has been shown to be enhanced by using legal receiver's feedback.
The present secret key based feedback scheme mainly focuses on producing key from the feedback and using this key
to protect part of the transmitted message.
However, this feedback scheme has been proved only optimal for several degraded cases.
The broadcast channel with mutual secrecy requirement (BC-MSR) is important as it constitutes the essence of physical layer security (PLS) in the down-link of the
wireless communication systems. In this paper, we investigate the feedback effects on the BC-MSR,
and propose two inner bounds and one outer bound on the secrecy capacity region of the BC-MSR with noiseless feedback. One inner bound is constructed according to the
already existing secret key based
feedback coding scheme for the wiretap channel, and the other is constructed by a hybrid coding scheme using feedback 
to generate not only keys protecting the transmitted messages
but also cooperative messages helping the receivers to improve their decoding performance.
The performance of the proposed feedback schemes and the gap between the inner and outer bounds are further explained via two examples.
\end{abstract}


\section{Introduction \label{secI}}

Besides reliability, introducing an additional secrecy criteria into a physically degraded
\footnote{The ``physically degraded'' indicates that the wiretapper's received signal is a degraded version of the legal receiver's.} broadcast channel,
Wyner \cite{Wy} first studied the secure transmission over the wiretap channel (WTC).
Later, on the basis of \cite{Wy}, Csisz$\acute{a}$r and K\"{o}rner \cite{CK} studied 
the WTC without the ``physically degraded'' assumption and with an additional common message available at both the legal receiver and the wiretapper.
The outstanding work \cite{Wy}-\cite{CK} reveals the reliability-security trade-off of the communication channels in the presence of a wiretapper.
The follow-up study of the WTC mainly focuses on the multi-user channel in the presence of a wiretapper (e.g. multiple-access wiretap channel \cite{tekin, dai1},
relay-eavesdropper channel \cite{relay1, relay2}, broadcast wiretap channel \cite{bwc,bwc1}, two way wiretap channel \cite{tw1, tw2}, etc.), and
the multi-terminal security, see \cite{LP}-\cite{dai2}.

In recent years, the effect of legal receiver's feedback on the PLS of communication channels attracts a lot of attention. 
For the WTC with noiseless feedback (WTC-NF), 
Ahlswede and Cai \cite{AC} pointed out that
to enhance the secrecy capacity of the WTC, the best use of the legal receiver's feedback channel output is to generate random bits from it
and use these bits as a key by the transmitter protecting part of the transmitted message. Using this secret key based feedback scheme,
Ahlswede and Cai \cite{AC} determined the secrecy capacity of the physically degraded WTC-NF, and 
the secrecy capacity of the general WTC-NF has not been determined yet.
On the basis of \cite{AC}, Ardestanizadeh et al. investigated the WTC with rate limited feedback \cite{AFJK}
where the legal receiver is free to use the noiseless feedback channel to send anything
as he wishes (up to a rate $R_{f}$). For the degraded case, they showed that the best choice of the legal receiver
is sending a key through the feedback channel, and if the legal receiver's channel output $Y_{1}$ is sent, the best use of it is to extract a key.
Later, Schaefer et al. \cite{bck} extended the work of \cite{AFJK}
to a broadcast situation, where two legitimate receivers of the broadcast channel independently sent their secret keys to the transmitter via two noiseless
feedback channels, and these keys help to increase the achievable secrecy rate region of the broadcast wiretap channel \cite{bwc}.
Cohen er al. \cite{cohen} generalized Ardestanizadeh et al.'s work \cite{AFJK} by
considering the WTC with noiseless feedback, and with
causal channel state information (CSI) at both the transmitter and the legitimate receiver. Cohen er al. \cite{cohen} showed that
the transmitted message can be protected by two keys, where one is generated from the noiseless feedback, and the other is generated by the causal CSI.
They further showed that these two keys increases the achievable secrecy rate of the WTC with rate limited feedback \cite{AFJK}.
Other related works in the WTC with noiseless feedback and CSI are investigated in \cite{dainew2}-\cite{dainew3}.
Here note that for the WTC-NF, the present literature (\cite{AC}-\cite{dainew3}) shows that the secrecy capacity is achieved only
for the degraded case, i.e.,  Ahlswede and Cai's secret key based feedback coding scheme \cite{AC}
is only optimal for the degraded channel models. Finding the optimal feedback coding scheme for the general channel models needs us to exploit
other uses of the feedback.

\begin{figure}[htb]
\centering
\includegraphics[scale=0.5]{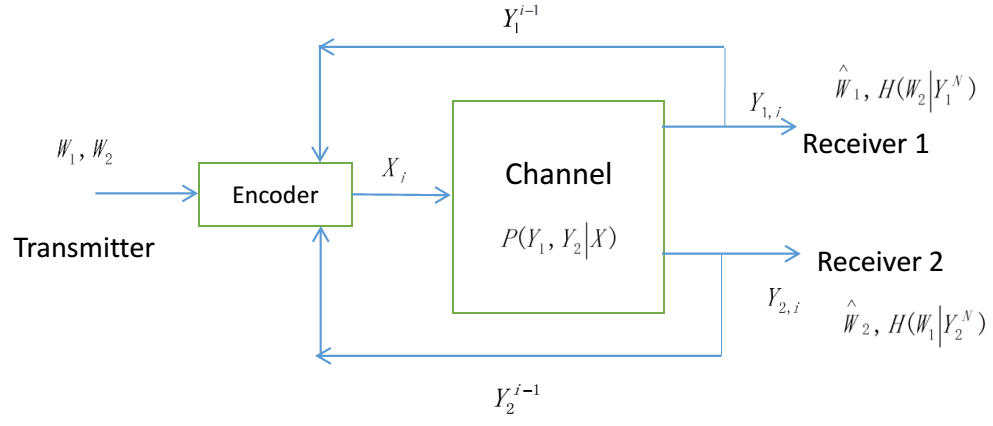}
\caption{The broadcast channel with noiseless feedback and mutual secrecy requirement}
\label{f1}
\end{figure}

The broadcast channel with mutual secrecy requirement (BC-MSR) is an important model for the PLS in the down-link of the wireless communication systems.
The already existing literature \cite{bc1, bc2} provides inner and outer bounds on the secrecy capacity region of BC-MSR.
To investigate the feedback effects on the BC-MSR (see Figure \ref{f1}), in this paper, two feedback strategies for the BC-MSR are proposed. One
is an extension of the already existing secret key based feedback scheme for the WTC \cite{AC}, and the other is a hybrid coding scheme
using the feedback to generate not only keys but also cooperative messages helping the receivers to improve their decoding performance.
Two inner bounds on the secrecy capacity region of the feedback model shown in Figure \ref{f1} are constructed with respect to
the proposed two feedback coding schemes. Moreover, for comparison, we also provide a corresponding outer bound.
These inner and outer bounds are further illustrated via
two examples (a Dueck type example and a Blackwell type example).

Now the remiander of this paper is organized as follows. Necessary mathematical background, the previous non-feedback coding scheme for the BC-MSR and
a generalized Wyner-Ziv coding scheme are provided in Section \ref{secII}.
Section \ref{secIII} is for the model formulation and the main results.
Section \ref{secIV} shows the details about the proof of the main results.
Two examples and numerical results
are shown in Section \ref{secV}, and a summary of this work is given in Section \ref{secVI}.

\section{Preliminaries}\label{secII}
\setcounter{equation}{0}

\subsection{Notations and Basic Lemmas}

\emph{Notations}: For the rest of this paper, the random variables (RVs), values and alphabets are written in uppercase letters, 
lowercase letters and calligraphic letters, respectively.
The random vectors and their values are denoted by a similar convention.
For example, $Y_{1}$ represents a RV, and $y_{1}$ represents a value in $\mathcal{Y}_{1}$. Similarly,
$Y_{1}^{N}$ represents a random $N$-vector $(Y_{1,1},...,Y_{1,N})$,
and $y_{1}^{N}=(y_{1,1},...,y_{1,N})$ represents a vector value in $\mathcal{Y}_{1}^{N}$
(the $N$-th Cartesian power of $\mathcal{Y}_{1}$).
In addition, for an event $X=x$, its probability is denoted by $P(x)$. In the remainder of this paper, the base of the $\log$ function is $2$.

An independent identically distributed (i.i.d.)
generated vector $x^{N}$ according to the probability $P(x)$ is $\epsilon$-typical if for all $x\in \mathcal{X}$,
$$|\frac{\pi_{x^{N}}(x)}{N}-P(x)|\leq \epsilon,$$
where $\pi_{x^{N}}(x)$ is the number of $x$ showing up in $x^{N}$.
The set composed of all typical vectors $x^{N}$ is called the strong typical set, and it is denoted by $T^{N}_{\epsilon}(P(x))$.
The following lemmas related with $T^{N}_{\epsilon}(P(x))$ will be used in the rest of this paper.

\begin{lemma}\label{L1}
\textbf{(Covering Lemma \cite{network})}: Let $X^{N}$ and $Y^{N}(l)$ ($l\in \mathcal{L}$ and $|\mathcal{L}|\geq 2^{NR}$) be i.i.d.
generated random vectors with respect to (w.r.t.) the probabilities $P(x)$ and $P(y)$, respectively. Here notice that $X^{N}$ is independent of $Y^{N}(l)$.
Then there exists $\nu>0$ satisfying the condition that 
\begin{eqnarray*}
\lim_{N\rightarrow\infty}P(\forall l\in \mathcal{L},\,\,(X^{N},Y^{N}(l))\notin T^{N}_{\nu}(P(x,y)))=0
\end{eqnarray*}
if $R>I(X;Y)+\varphi(\nu)$, where $\varphi(\nu)\rightarrow 0$ as $\nu\rightarrow 0$.
\end{lemma}

\begin{lemma}\label{L2}
\textbf{(Packing Lemma \cite{network})}: Let $X^{N}$ and $Y^{N}(l)$ ($l\in \mathcal{L}$ and $|\mathcal{L}|\leq 2^{NR}$) be i.i.d.
generated random vectors w.r.t. the probabilities $P(x)$ and $P(y)$, respectively. Here notice that $X^{N}$ is independent of $Y^{N}(l)$.
Then there exists $\nu>0$ satisfying the condition that
\begin{eqnarray*}
\lim_{N\rightarrow\infty}P(\exists l\in \mathcal{L} \,\,s.t.\,\,(X^{N},Y^{N}(l))\in T^{N}_{\nu}(P(x,y)))=0
\end{eqnarray*}
if $R<I(X;Y)-\varphi(\nu)$, where $\varphi(\nu)\rightarrow 0$ as $\nu\rightarrow 0$.
\end{lemma}

\begin{lemma}\label{L4}
\textbf{(Balanced coloring lemma \cite[p. 260]{AC})}: For any $\epsilon, \delta>0$ and sufficiently large $N$,
let $Q^{N}$, $U_{1}^{N}$, $U_{2}^{N}$, $V_{2}^{N}$, $Y_{1}^{N}$ and $Y_{2}^{N}$ be i.i.d. generated random vectors
respectively w.r.t. the probabilities $P(q)$, $P(u_{1})$, $P(u_{2})$, $P(v_{2})$, $P(y_{1})$ and $P(y_{2})$. Given $y_{2}^{N}$,
$P(q)$, $P(u_{1})$, $P(u_{2})$ and $P(v_{2})$, let
$T_{P(y_{1}|y_{2},q,u_{1},u_{2},v_{2})}^{N}(y_{2}^{N},q^{N},u_{1}^{N},u_{2}^{N},v_{2}^{N})$ be the conditional strong typical set composed of all $y_{1}^{N}$
satisfying the fact that
$(y_{1}^{N},y_{2}^{N},q^{N},u_{1}^{N},u_{2}^{N},v_{2}^{N})$ are jointly typical. In addition, for
$\gamma<|T_{P(y_{1}|y_{2},q,u_{1},u_{2},v_{2})}^{N}(y_{2}^{N},q^{N},u_{1}^{N},u_{2}^{N},v_{2}^{N})|$,
let $\phi$ be a $\gamma$-coloring
$$\phi: T^{N}_{\epsilon}(P(y_{1}))\rightarrow \{1,2,..,\gamma\},$$
and $\phi^{-1}(k)$ ($k\in \{1,2,..,\gamma\}$) be a set composed of all $y_{1}^{N}$ such that
$\phi(y_{1}^{N})=k$ and $y_{1}^{N}\in T_{P(y_{1}|y_{2},q,u_{1},u_{2},v_{2})}^{N}(y_{2}^{N},q^{N},u_{1}^{N},u_{2}^{N},v_{2}^{N})$.
Then we have
\begin{equation}\label{aaa1}
|\phi^{-1}(k)|\leq \frac{|T_{P(y_{1}|y_{2},q,u_{1},u_{2},v_{2})}^{N}(y_{2}^{N},q^{N},u_{1}^{N},u_{2}^{N},v_{2}^{N})|(1+\delta)}{\gamma},
\end{equation}
where $k\in \{1,2,..,\gamma\}$.
\end{lemma}
\begin{remark}\label{dd1}
From Lemma \ref{L4}, it is easy to see that there are at least
\begin{equation}\label{aaa2}
\frac{|T_{P(y_{1}|y_{2},q,u_{1},u_{2},v_{2})}^{N}(y_{2}^{N},q^{N},u_{1}^{N},u_{2}^{N},v_{2}^{N})|}
{\frac{|T_{P(y_{1}|y_{2},q,u_{1},u_{2},v_{2})}^{N}(y_{2}^{N},q^{N},u_{1}^{N},u_{2}^{N},v_{2}^{N})|(1+\delta)}{\gamma}}=\frac{\gamma}{1+\delta}
\end{equation}
colors and at most $\gamma$ colors
mapped by $T_{P(y_{1}|y_{2},q,u_{1},u_{2},v_{2})}^{N}(y_{2}^{N},q^{N},u_{1}^{N},u_{2}^{N},v_{2}^{N})$.
Letting $\gamma=|T_{P(y_{1}|y_{2},q,u_{1},u_{2},v_{2})}^{N}(y_{2}^{N},q^{N},u_{1}^{N},u_{2}^{N},v_{2}^{N})|$ and applying the properties of the conditional
strong typical set \cite{network}, we see that
\begin{eqnarray}\label{zhenni1}
&&\gamma=|T_{P(y_{1}|y_{2},q,u_{1},u_{2},v_{2})}^{N}(y_{2}^{N},q^{N},u_{1}^{N},u_{2}^{N},v_{2}^{N})|\geq (1-\epsilon_{1})2^{N(1-\epsilon_{2})
H(Y_{1}|Y_{2},Q,U_{1},U_{2},V_{2})},
\end{eqnarray}
\end{remark}
where $\epsilon_{1}$ and $\epsilon_{2}$ tend to zero while $N\rightarrow \infty$.

\subsection{Non-feedback coding scheme for the BC-MSR}\label{aa1}

For the model of Figure \ref{f1} without feedback, a hybrid coding scheme combining
Marton's binning technique for the general broadcast channel \cite{ma} with the random binning technique for the wiretap channel \cite{Wy}
is proposed in \cite{bc1, bc2}. In this subsection, we review this hybrid coding scheme.

\emph{Definitions}:

The message $W_{j}$ ($j=1,2$) is conveyed to Receiver $j$, and it is uniformly drawn from the set $\{1,...,2^{NR_{j}}\}$.
The randomly generated $W^{'}_{j}$, which is used for confusing the illegal receiver \footnote{The idea of using random
messages to confuse the wiretapper is exactly the same as the random binning technique used in Wyner's wiretap channel \cite{Wy}, where
this randomly produced message is analogous to the randomly chosen bin index used in the random binning scheme.},
is uniformly drawn from the set $\{1,...,2^{NR_{j}^{'}}\}$, i.e.,
$Pr\{W^{'}_{j}=i\}=2^{-NR_{j}^{'}}$, where $i\in \{1,...,2^{NR_{j}^{'}}\}$.
Moreover, similar to the coding scheme in Marton's achievable region for the broadcast channel \cite{ma},
the message $W^{''}_{j}$, which enables its codeword $U^{N}_{j}$ to be jointly typical with other codewords,
chooses values from the set $\{1,...,2^{NR_{j}^{''}}\}$.

\emph{Code construction}:

First, randomly generate $2^{NR_{0}}$ i.i.d. $Q^{N}$ w.r.t. $P(q)$, and label them as
$q^{N}(w_{0})$, where $w_{0}\in \{1,2,...,2^{NR_{0}}\}$.
Then, for each possible value of $q^{N}$, randomly generate $2^{N(R_{j}+R_{j}^{'}+R_{j}^{''})}$ i.i.d. $U^{N}_{j}$ w.r.t. $P(u_{j}|q)$,
and label them as $u^{N}_{j}(w_{j},w^{'}_{j},w^{''}_{j})$, where
$w_{j}\in \{1,2,...,2^{NR_{j}}\}$, $w^{'}_{j}\in \{1,2,...,2^{NR^{'}_{j}}\}$ and
$w^{''}_{j}\in \{1,2,...,2^{NR^{''}_{j}}\}$.
Finally, for each possible value of $q^{N}$, $u^{N}_{1}$ and $u^{N}_{2}$,
the channel input $x^{N}$ is i.i.d. generated w.r.t. $P(x|q,u_{1},u_{2})$.

\emph{Encoding procedure}:

The transmitter selects $q^{N}(w_{0})$,
$u^{N}_{1}(w_{1},w^{'}_{1},w^{''}_{1})$ and $u^{N}_{2}(w_{2},w^{'}_{2},w^{''}_{2})$ to transmit.
Here notice that $w_{0}$, $w^{'}_{1}$ and $w^{'}_{2}$ are randomly chosen from the sets $\{1,2,...,2^{NR_{0}}\}$, $\{1,2,...,2^{NR^{'}_{1}}\}$ and
$\{1,2,...,2^{NR^{'}_{2}}\}$, respectively, and the indexes $w^{''}_{1}$ and $w^{''}_{2}$ are chosen by finding a pair of
$(u^{N}_{1}(w_{1},w^{'}_{1},w^{''}_{1}),u^{N}_{2}(w_{2},w^{'}_{2},w^{''}_{2}))$
satisfying the condition that given $q^{N}(w_{0})$, $(u^{N}_{1},u^{N}_{2},q^{N}(w_{0}))$ are jointly typical. If multiple pairs exist,
choose the pair with the smallest indexes; if no such pair exists, proclaim an encoding error. On the basis of the covering lemma (see Lemma \ref{L1}),
this kind of encoding error tends to zero if
\begin{eqnarray}\label{vp1}
&&R^{''}_{1}+R^{''}_{2}\geq I(U_{1};U_{2}|Q).
\end{eqnarray}

\emph{Decoding procedure}:

First, Receiver $j$ ($j=1,2$) chooses a unique $q^{N}$
jointly typical with $y_{j}^{N}$. if more than one or no such $q^{N}$ exists, declare
an decoding error.
From packing lemma (see Lemma \ref{L2}), this kind of decoding error tends to zero if
\begin{eqnarray}\label{vp2}
&&R_{0}\leq I(Y_{j};Q).
\end{eqnarray}
After decoding $q^{N}$, Receiver $j$ seeks a unique $u^{N}_{j}$ satisfying the condition that $(u^{N}_{j},q^{N},y_{j}^{N})$ are jointly
typical. From the packing lemma, this kind of decoding error tends to zero if
\begin{eqnarray}\label{vp3}
&&R_{j}+R^{'}_{j}+R^{''}_{j}\leq I(Y_{j};U_{j}|Q).
\end{eqnarray}
Once $u^{N}_{j}$ is decoded, Receiver $j$ extracts $w_{j}$ in it.

\emph{Equivocation analysis}:

The Receiver $2$'s equivocation rate $\Delta_{1}$, which is denoted by $\Delta_{1}=\frac{1}{N}H(W_{1}|Y_{2}^{N})$, follows that
\begin{eqnarray}\label{vp4}
&&\Delta_{1}=\frac{1}{N}H(W_{1}|Y_{2}^{N})\geq \frac{1}{N}H(W_{1}|Y_{2}^{N},Q^{N},U_{2}^{N})\nonumber\\
&&=\frac{1}{N}(H(W_{1},Y_{2}^{N},Q^{N},U_{2}^{N})-H(Y_{2}^{N},Q^{N},U_{2}^{N}))\nonumber\\
&&=\frac{1}{N}(H(U_{1}^{N},W_{1},Y_{2}^{N},Q^{N},U_{2}^{N})-H(U_{1}^{N}|W_{1},Y_{2}^{N},Q^{N},U_{2}^{N})-H(Y_{2}^{N},Q^{N},U_{2}^{N}))\nonumber\\
&&\stackrel{(a)}=\frac{1}{N}(H(Y_{2}^{N}|Q^{N},U_{2}^{N},U_{1}^{N})
+H(U_{1}^{N}|Q^{N},U_{2}^{N})-H(U_{1}^{N}|W_{1},Y_{2}^{N},Q^{N},U_{2}^{N})-H(Y_{2}^{N}|Q^{N},U_{2}^{N}))\nonumber\\
&&=\frac{1}{N}(H(U_{1}^{N}|Q^{N})-I(U_{1}^{N};U_{2}^{N}|Q^{N})-H(U_{1}^{N}|W_{1},Y_{2}^{N},Q^{N},U_{2}^{N})
-I(Y_{2}^{N};U_{1}^{N}|Q^{N},U_{2}^{N}))\nonumber\\
&&\stackrel{(b)}=\frac{1}{N}(N(R_{1}+R_{1}^{'}+R_{1}^{''})-NI(U_{1};U_{2}|Q)-NI(Y_{2};U_{1}|Q,U_{2})-H(U_{1}^{N}|W_{1},Y_{2}^{N},Q^{N},U_{2}^{N}))\nonumber\\
&&\stackrel{(c)}\geq R_{1}+R_{1}^{'}+R_{1}^{''}-I(U_{1};U_{2}|Q)-I(Y_{2};U_{1}|Q,U_{2})-\delta(\epsilon_{1}),
\end{eqnarray}
where (a) follows from $H(W_{1}|U_{1}^{N})=0$, (b) follows from the generation of $Q^{n}$, $U_{1}^{n}$, $U_{2}^{n}$ and
the channel is memoryless, and (c) follows from
that given $\tilde{w}_{1}$, $q^{N}$, $u_{2}^{N}$ and $y_{2}^{n}$, Receiver $2$ tries 
to find out only one $u_{1}^{n}$ that is jointly typical with $y_{2}^{n}$, $q^{n}$, $u_{2}^{n}$, and
implied by the packing lemma, we see that
Receiver $2$'s decoding error tends to zero if
\begin{eqnarray}\label{vp5}
&&R_{1}+R_{1}^{'}\leq I(Y_{2};U_{1}|Q,U_{2}),
\end{eqnarray}
then applying Fano's lemma, $\frac{1}{N}H(U_{1}^{N}|W_{1},Y_{2}^{N},Q^{N},U_{2}^{N})\leq \epsilon_{1}$ is obtained,
where $\epsilon_{1}\rightarrow 0$ while $N\rightarrow \infty$.
From (\ref{vp4}), we can conclude that $\Delta_{1}\geq R_{1}-\epsilon$ if
\begin{eqnarray}\label{vp6}
&&R_{1}^{'}+R_{1}^{''}\geq I(U_{1};U_{2}|Q)+I(Y_{2};U_{1}|Q,U_{2}).
\end{eqnarray}
Analogously, we can conclude that $\Delta_{2}\geq R_{2}-\epsilon$ if
\begin{eqnarray}\label{vp7}
&&R_{2}^{'}+R_{2}^{''}\geq I(U_{1};U_{2}|Q)+I(Y_{1};U_{2}|Q,U_{1}),
\end{eqnarray}
and
\begin{eqnarray}\label{vp8}
&&R_{2}+R_{2}^{'}\leq I(Y_{1};U_{2}|Q,U_{1}).
\end{eqnarray}
From (\ref{vp1}), (\ref{vp3}), (\ref{vp5}), (\ref{vp6}), (\ref{vp7}) and (\ref{vp8}), the achievable secrecy rate region
$\mathcal{C}_{bc-msr}$ for the BC-MSR \cite{bc1} is obtained, and it is given by
\begin{eqnarray*}
&&\mathcal{C}_{bc-msr}=\{(R_{1}, R_{2}): 0\leq R_{1}\leq I(Y_{1};U_{1}|Q)-I(U_{1};U_{2}|Q)-I(Y_{2};U_{1}|Q,U_{2})\\
&&0\leq R_{2}\leq I(Y_{2};U_{2}|Q)-I(U_{1};U_{2}|Q)-I(Y_{1};U_{2}|Q,U_{1})\}.
\end{eqnarray*}

Combining the above coding scheme for $\mathcal{C}_{bc-msr}$ with
the already existing secret key based feedback scheme for the WTC \cite{AC}, it is not difficult to propose a secret key based feedback coding scheme
for the BC-MSR, which will be shown in the next section.

\subsection{The Generalized Wyner-Ziv Coding Scheme for the Distributed Source Coding with Side Information}\label{aa2}

\begin{figure}[htb]
\centering
\includegraphics[scale=0.5]{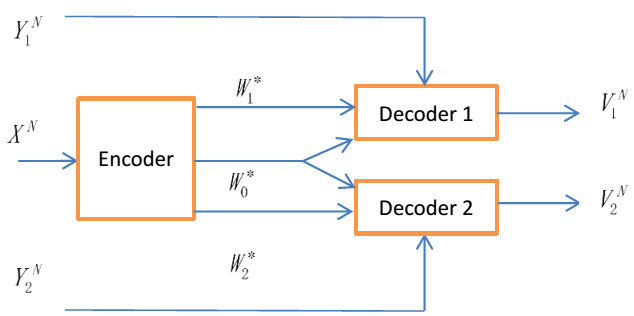}
\caption{The distributed source coding with side information}
\label{jinsha1}
\end{figure}

In this subsection, we review the generalized Wyner-Ziv coding scheme for the distributed source coding with side information \cite{kramer}.
For the distributed source coding with side information shown in Figure \ref{jinsha1},
the source $X^{N}$ is correlated with the side information $Y_{1}^{N}$ and $Y_{2}^{N}$, and they are
i.i.d. generated according to the probability $P(x,y_{1},y_{2})$. Using an encoding function
$\phi: \mathcal{X}^{N}\rightarrow \{1,2,...,2^{NR_{0}}\}\times \{1,2,...,2^{NR_{1}}\}\times \{1,2,...,2^{NR_{2}}\}$,
the transmitter compresses $X^{N}$ into three indexes $W^{*}_{0}$, $W^{*}_{1}$ and $W^{*}_{2}$ respectively choosing values from the sets $\{1,2,...,2^{NR_{0}}\}$,
$\{1,2,...,2^{NR_{1}}\}$ and $\{1,2,...,2^{NR_{2}}\}$.
The indexes $W^{*}_{0}$, $W^{*}_{j}$ ($j=1,2$) together with the side information $Y_{j}^{N}$ are available at Receiver $j$.
Receiver $j$ generates a reconstruction sequence $\hat{V}_{j}^{N}=\varphi(W^{*}_{0},W^{*}_{j},Y_{j}^{N})$ by applying a reconstruction function
$\varphi_{j}: \{1,2,...,2^{NR_{0}}\}\times \{1,2,...,2^{NR_{j}}\}\times \mathcal{Y}_{j}^{N}\rightarrow \mathcal{V}_{j}^{N}$ to the indexes $W^{*}_{0}$, $W^{*}_{j}$ and the
side information $Y_{j}^{N}$. The goal of the communication is that the reconstruction sequence $\hat{V}_{j}^{N}$ is jointly typical with the source
$X^{N}$ according to the probability $P(v_{j}|x)\times P(x)$.

A rate triplet $(R_{0},R_{1},R_{2})$ is said to be achievable if for any $\epsilon>0$, there exists
a sequence of encoding and reconstruction functions $(\phi,\varphi_{1},\varphi_{2})$ such that
\begin{eqnarray}\label{kong1-sg1}
&&Pr\{(X^{N},\hat{V}_{j}^{N})\notin T^{N}_{\epsilon}(P(x,v_{j}))\}\rightarrow 0
\end{eqnarray}
as $N\rightarrow \infty$.
The following generalized Wyner-Ziv Theorem \cite{kramer} provides an achievable region $\mathcal{R}_{inner}$
consisting of achievable rate triplets $(R_{0},R_{1},R_{2})$
for this distributed source coding with side information problem.

\begin{theorem}\label{T2-xx}
(Generalized Wyner-Ziv Theorem): For the distributed source coding with side information, an achievable rate region $\mathcal{R}_{inner}$ is given by
\begin{eqnarray}\label{trans1x}
&&\mathcal{R}_{inner}=\{(R_{0}, R_{1}, R_{2}): R_{0}+R_{1}\geq I(X;V_{0},V_{1}|Y_{1})\nonumber\\
&&R_{0}+R_{2}\geq I(X;V_{0},V_{2}|Y_{2})\nonumber\\
&&R_{0}+R_{1}+R_{2}\geq I(X;V_{1}|Y_{1},V_{0})+I(X;V_{2}|Y_{2},V_{0})+\max_{j\in\{1,2\}}I(X;V_{0}|Y_{j})\},
\end{eqnarray}
where $(V_{0},V_{1},V_{2})\rightarrow X\rightarrow (Y_{1},Y_{2})$.
\end{theorem}

\emph{Achievable coding scheme for Theorem \ref{T2-xx}}:

\begin{itemize}

\item \textbf{Definitions}:
The index $w^{*}_{0}$ chooses values from the set $\{1,2,...,2^{NR_{0}}\}$, and divide $w^{*}_{0}$ 
into three sub-indexes $w^{*}_{0,0}$, $w^{*}_{0,1}$ and $w^{*}_{0,2}$, where
each sub-index $w^{*}_{0,i}$ ($i\in\{0,1,2\}$) chooses values from the set $\{1,2,...,2^{NR_{0,i}}\}$ and $R_{0,0}+R_{0,1}+R_{0,2}=R_{0}$.
The index $w^{*}_{1}$ chooses values from $\{1,2,...,2^{NR_{1}}\}$, and divide $w^{*}_{1}$ into two sub-indexes $w^{*}_{1,0}$ and $w^{*}_{1,1}$, where
each sub-index $w^{*}_{1,j}$ ($j\in\{0,1\}$) chooses values from $\{1,2,...,2^{NR_{1,j}}\}$ and $R_{1,0}+R_{1,1}=R_{1}$.
Similarly, the index $w^{*}_{2}$ chooses values from $\{1,2,...,2^{NR_{2}}\}$, and divide $w^{*}_{2}$ into two sub-indexes $w^{*}_{2,0}$ and $w^{*}_{2,2}$, where
each sub-index $w^{*}_{2,l}$ ($l\in\{0,2\}$) chooses values from $\{1,2,...,2^{NR_{2,l}}\}$ and $R_{2,0}+R_{2,2}=R_{2}$.
Define $k_{1,0}$, $k_{2,0}$, $k_{1}$ and $k_{2}$ as auxiliary indexes respectively taking values in
$\{1,2,...,2^{N(R^{'}_{0}-R_{1,0})}\}$, $\{1,2,...,2^{N(R^{'}_{0}-R_{2,0})}\}$, $\{1,2,...,2^{NR^{'}_{1}}\}$ and $\{1,2,...,2^{NR^{'}_{2}}\}$.

\item \textbf{Code-book generation}:
There are two different ways to generate the sequence $v_{0}^{N}$. The first way is to generate $2^{N(R_{0,0}+R_{0}^{'})}$ i.i.d. sequences
$v_{0}^{N}(1;w^{*}_{0,0},w^{*}_{1,0},k_{1,0})$ with respect to (w.r.t.) the probability $P(v_{0})$. The second way is to generate $2^{N(R_{0,0}+R_{0}^{'})}$ i.i.d. sequences
$v_{0}^{N}(2;w^{*}_{0,0},w^{*}_{2,0},k_{2,0})$ w.r.t. the probability $P(v_{0})$.
Here note that $v_{0}^{N}(j;w^{*}_{0,0},w^{*}_{j,0},k_{j,0})$ ($j\in\{1,2\}$) is intended to be decoded by Receiver $j$.
Then, generate $2^{N(R_{0,1}+R_{1,1}+R_{1}^{'})}$ i.i.d. sequences
$v_{1}^{N}(w^{*}_{0,1},w^{*}_{1,1},k_{1})$ w.r.t. the probability $P(v_{1})$, and generate $2^{N(R_{0,2}+R_{2,2}+R_{2}^{'})}$ i.i.d. sequences
$v_{2}^{N}(w^{*}_{0,2},w^{*}_{2,2},k_{2})$ w.r.t. the probability $P(v_{2})$.

\item \textbf{Encoding}:
Given a source $x^{N}$, the encoder seeks a pair of sequences
$(v_{0}^{N}(j;\tilde{w}^{*}_{0,0},\tilde{w}^{*}_{j,0},\tilde{k}_{j,0}),\\v_{j}^{N}(\tilde{w}^{*}_{0,j},\tilde{w}^{*}_{j,j},\tilde{k}_{j}))$ ($j\in\{1,2\}$)
satisfying the condition that $(x^{N},v_{0}^{N}(j;\tilde{w}^{*}_{0,0},\tilde{w}^{*}_{j,0},\tilde{k}_{j,0}),v_{j}^{N}(\tilde{w}^{*}_{0,j},\tilde{w}^{*}_{j,j},\tilde{k}_{j}))$
are jointly typical.
If there is more than one such pair, randomly choose one. If there is no such pair, declare an encoding error.
On the basis of the covering lemma (see \ref{L1}), the encoding error tends to zero if
\begin{eqnarray}\label{vp1-1}
&&R^{'}_{0}+R_{0,0}\geq I(X;V_{0}),
\end{eqnarray}
and
\begin{eqnarray}\label{vp1-2}
&&R^{'}_{j}+R_{0,j}+R_{j,j}\geq I(V_{j};X,V_{0}).
\end{eqnarray}
Once the sequences $v_{0}^{N}(j;\tilde{w}^{*}_{0,0},\tilde{w}^{*}_{j,0},\tilde{k}_{j,0})$ and $v_{j}^{N}(\tilde{w}^{*}_{0,j},\tilde{w}^{*}_{j,j},\tilde{k}_{j})$
are chosen for $j\in\{1,2\}$, the encoder sends the index $w^{*}_{0}=(\tilde{w}^{*}_{0,0},\tilde{w}^{*}_{0,1},\tilde{w}^{*}_{0,2})$ to both receivers,
sends $w^{*}_{1}=(\tilde{w}^{*}_{1,0},\tilde{w}^{*}_{1,1})$ to Receiver $1$ only, and sends $w^{*}_{2}=(\tilde{w}^{*}_{2,0},\tilde{w}^{*}_{2,2})$ to Receiver $2$ only.

\item \textbf{Decoding}:
Upon receiving the indexes $w^{*}_{0}$ and $w^{*}_{j}$ ($j\in\{1,2\}$), Receiver $j$ parses the common index $w^{*}_{0}$ as
$(\tilde{w}^{*}_{0,0},\tilde{w}^{*}_{0,1},\tilde{w}^{*}_{0,2})$, and its private index $w^{*}_{j}$ as
$(\tilde{w}^{*}_{j,0},\tilde{w}^{*}_{j,j})$. Then given the side information $y_{j}^{N}$ and $\tilde{w}^{*}_{0,0}$, $\tilde{w}^{*}_{0,1}$, $\tilde{w}^{*}_{0,2}$,
$\tilde{w}^{*}_{j,0}$, $\tilde{w}^{*}_{j,j}$,
Receiver $j$ seeks a unique pair of
$(v_{0}^{N}(j;\tilde{w}^{*}_{0,0},\tilde{w}^{*}_{j,0},\hat{k}_{j,0}),v_{j}^{N}(\tilde{w}^{*}_{0,j},\tilde{w}^{*}_{j,j},\hat{k}_{j}))$ satisfying the condition that
$(v_{0}^{N}(j;\tilde{w}^{*}_{0,0},\tilde{w}^{*}_{j,0},\hat{k}_{j,0}),\\ v_{j}^{N}(\tilde{w}^{*}_{0,j},\tilde{w}^{*}_{j,j},\hat{k}_{j}),y_{j}^{N})$ are jointly typical.
If there is no or more than one pair, declare an decoding error.
On the basis of the packing lemma (see \ref{L2}), Receiver $j$'s decoding error tends to zero if
\begin{eqnarray}\label{vp1-3}
&&R^{'}_{j}\leq I(V_{j};V_{0},Y_{j}),
\end{eqnarray}
and
\begin{eqnarray}\label{vp1-4}
&&R^{'}_{0}-R_{j,0}+R^{'}_{j}\leq I(V_{0};Y_{j})+I(V_{j};V_{0},Y_{j}).
\end{eqnarray}
Once Receiver $j$ finds such unique pair of $(v_{0}^{N}(j;\tilde{w}^{*}_{0,0},\tilde{w}^{*}_{j,0},\hat{k}_{j,0}),
v_{j}^{N}(\tilde{w}^{*}_{0,j},\tilde{w}^{*}_{j,j},\hat{k}_{j}))$,
he generates the re-construction sequence $\hat{V}_{j}^{N}=v_{j}^{N}(\tilde{w}^{*}_{0,j},\tilde{w}^{*}_{j,j},\hat{k}_{j})$.

\item Using the fact that $R_{0,0}+R_{0,1}+R_{0,2}=R_{0}$, $R_{1,0}+R_{1,1}=R_{1}$, $R_{2,0}+R_{2,2}=R_{2}$, and
applying Fourier-Motzkin elimination to remove $R^{'}_{0}$, $R^{'}_{1}$ and $R^{'}_{2}$ from
(\ref{vp1-1}), (\ref{vp1-2}), (\ref{vp1-3}) and (\ref{vp1-4}), Theorem \ref{T2-xx} is proved.

\end{itemize}

Here note that the generalized Wyner-Ziv coding scheme described above indicates that in a broadcast channel, each receiver's channel output can be viewed
as side information helping the receiver to decode an estimation of the channel input, and this estimation of the channel input helps the receiver to
improve his decoding performance. Motivated by this generalized Wyner-Ziv coding scheme, in the next section, a hybrid feedback strategy
for the BC-MSR is proposed, which combines the already existing secret key based feedback scheme for the WTC \cite{AC} and the generalized Wyner-Ziv coding scheme
with the previous non-feedback coding scheme for the BC-MSR described in Subsection \ref{aa1}.

\section{Problem Formulation and Main Results}\label{secIII}
\setcounter{equation}{0}

The model of BC-MSR consists of one input $x^{N}$, two outputs $y_{1}^{N}$, $y_{2}^{N}$, and satisfies
\begin{equation}\label{shitu1}
P(y_{1}^{N},y_{2}^{N}|x^{N})=\prod_{i=1}^{n}P(y_{1,i},y_{2,i}|x_{i}),
\end{equation}
where $x_{i}\in \mathcal{X}$, $y_{1,i}\in \mathcal{Y}_{1}$ and $y_{2,i}\in \mathcal{Y}_{2}$.

Let $W_{1}$ and $W_{2}$ be the transmission messages, and their values respectively belong to
the alphabets $\mathcal{W}_{1}=\{1,2,...,M_{1}\}$ and $\mathcal{W}_{2}=\{1,2,...,M_{2}\}$. In addition,
$Pr\{W_{1}=i\}=\frac{1}{M_{1}}$ for $i\in \mathcal{W}_{1}$, and $Pr\{W_{2}=j\}=\frac{1}{M_{2}}$ for $j\in \mathcal{W}_{2}$.
Using feedback, the transmitter produces the time-$t$ channel input $X_{t}$ as a function of the messages $W_{1}$, $W_{2}$ and of
the previously received channel outputs $Y_{1,1}$,...,$Y_{1,t-1}$ and $Y_{2,1}$,...,$Y_{2,t-1}$, i.e.,
\begin{equation}\label{e202}
X_{t}=f_{t}(W_{1},W_{2},Y_{1}^{t-1},Y_{2}^{t-1})
\end{equation}
for some stochastic encoding function $f_{t}$ ($1\leq t\leq N$).

After $N$ channel uses, Receiver $j$ ($j=1,2$) decodes $W_{j}$. Namely, Receiver $j$ generates the guess
$$\hat{W}_{j}=\psi_{j}(Y_{j}^{N}),$$
where $\psi_{j}$ is Receiver $j$'s decoding function.
Receiver $j$'s average decoding error probability is denoted by
\begin{equation}\label{e204}
P_{e,j}=\frac{1}{M_{j}}\sum_{w_{j}\in \mathcal{W}_{j}}Pr\{\psi_{j}(y_{j}^{N})\neq w_{j}|w_{j}\,\,\mbox{sent}\}.
\end{equation}
Receiver $2$'s equivocation rate of the message $W_{1}$ is formulated as
\begin{equation}\label{e205}
\Delta_{1}=\frac{1}{N}H(W_{1}|Y_{2}^{N}).
\end{equation}
Analogously, Receiver $1$'s equivocation rate oft the message $W_{2}$ is formulated as
\begin{equation}\label{e205.ce}
\Delta_{2}=\frac{1}{N}H(W_{2}|Y_{1}^{N}).
\end{equation}

Define an achievable secrecy rate pair $(R_{1},R_{2})$ as below.
Given two positive numbers $R_{1}$ and $R_{2}$, if for arbitrarily small $\epsilon$,
there exist one channel encoder and two channel decoders
with parameters $M_{1}$, $M_{2}$, $N$, $\Delta_{1}$, $\Delta_{2}$, $P_{e,1}$ and $P_{e,2}$ satisfying
\begin{eqnarray}\label{e205}
&&\frac{\log M_{1}}{N}\geq R_{1}-\epsilon, \label{a1}\\
&&\frac{\log M_{2}}{N}\geq R_{2}-\epsilon, \label{a2}\\
&&\Delta_{1}\geq R_{1}-\epsilon, \label{a3}\\
&&\Delta_{2}\geq R_{2}-\epsilon,\label{a4}\\
&&P_{e,1}\leq \epsilon, \,P_{e,2}\leq \epsilon,\label{a5}
\end{eqnarray}
the pair $(R_{1},R_{2})$ is called an achievable secrecy rate pair.
The secrecy capacity region $\mathcal{C}^{f}_{s}$ consists of all achievable secrecy rate pairs.
We first propose a hybrid inner bound
$\mathcal{C}^{f-in-2}_{s}$
on $\mathcal{C}^{f}_{s}$. The feedback channel outputs $Y_{1}^{i-1}$ and $Y_{2}^{i-1}$ are not only used to generate secret keys protecting part of the messages,
but also used to produce cooperative messages represented by $V_{0}$, $V_{1}$ and $V_{2}$ helping the receivers to improve their decoding performances.
The inner bound $\mathcal{C}^{f-in-2}_{s}$
is provided in the following Theorem \ref{T2}.

\begin{theorem}\label{T2}
$\mathcal{C}^{f-in-2}_{s}\subseteq \mathcal{C}^{f}_{s}$,
where
\begin{eqnarray*}
&&\mathcal{C}^{f-in-2}_{s}=\{(R_{1}, R_{2}): 0\leq R_{1}\leq \min\{
[I(U_{1};Y_{1},V_{1}|Q)-I(U_{1};U_{2}|Q)-I(U_{1};Y_{2},V_{2}|Q,U_{2})]^{+}\\
&&+H(Y_{1}|Q,U_{1},U_{2},Y_{2},V_{2}),I(U_{1};Y_{1},V_{1}|Q)\},\\
&&0\leq R_{2}\leq \min\{
[I(U_{2};Y_{2},V_{2}|Q)-I(U_{1};U_{2}|Q)-I(U_{2};Y_{1},V_{1}|Q,U_{1})]^{+}+H(Y_{2}|Q,U_{1},U_{2},Y_{1},V_{1}),\\
&&I(U_{2};Y_{2},V_{2}|Q)\},\\
&&0\leq R_{1}\leq \min\{I(Q;Y_{1},V_{1}),I(Q;Y_{2},V_{2})\}+I(U_{1};Y_{1},V_{1}|Q)-I(V_{0},V_{1};Q,U_{1},U_{2},\tilde{Y}|Y_{1}),\\
&&0\leq R_{2}\leq \min\{I(Q;Y_{1},V_{1}),I(Q;Y_{2},V_{2})\}+I(U_{2};Y_{2},V_{2}|Q)-I(V_{0},V_{2};Q,U_{1},U_{2},\tilde{Y}|Y_{2}),\\
&&0\leq R_{1}+R_{2}\leq \min\{I(Q;Y_{1},V_{1}),I(Q;Y_{2},V_{2})\}+I(U_{1};Y_{1},V_{1}|Q)+I(U_{2};Y_{2},V_{2}|Q)\\
&&-I(U_{1};U_{2}|Q)
-I(V_{1};Q,U_{1},U_{2},\tilde{Y}|Y_{1},V_{0})-I(V_{2};Q,U_{1},U_{2},\tilde{Y}|Y_{2},V_{0})-\\
&&\max\{I(V_{0};Q,U_{1},U_{2},\tilde{Y}|Y_{1}),I(V_{0};Q,U_{1},U_{2},\tilde{Y}|Y_{2})\}\},
\end{eqnarray*}
$\tilde{Y}=(Y_{1},Y_{2})$ and the joint distribution is denoted by
\begin{eqnarray}\label{trans2.bg1}
&&P(q,u_{1},u_{2},v_{0},v_{1},v_{2},x,y_{1},y_{2})\nonumber\\
&&=P(v_{0},v_{1},v_{2}|q,u_{1},u_{2},y_{1},y_{2})P(y_{1},y_{2}|x)P(x|u_{1},u_{2})P(u_{1},u_{2}|q)P(q).
\end{eqnarray}
\end{theorem}
\begin{IEEEproof}
The coding scheme achieving the inner bound $\mathcal{C}^{f-in-2}_{s}$ combines the already existing secret key based feedback scheme for the WTC \cite{AC}
and the previous non-feedback coding scheme for the BC-MSR
with the generalized
Wyner-Ziv scheme described in Section \ref{secII}, and it can be briefly illustrated as follows.
\begin{itemize}
\item \emph{Encoding}: The transmission is through $n$ blocks. First, similar to the secret key based feedback scheme for the WTC \cite{AC}, in each block,
split the transmitted message $w_{i}$ ($i\in\{1,2\}$) into two parts, i.e., $w_{i}=(w_{i,1},w_{i,2})$. The sub-message $w_{i,1}$ is encoded exactly the same
as that in the non-feedback coding scheme for BC-MSR (see Section \ref{secII}), and $w_{i,2}$ is encrypted by a key produced by the feedback channel output
$y_{i}^{N}$ of the previous block.
Then, compress the encoded sequences $u_{1}^{N}$,
$u_{2}^{N}$, $q^{N}$ and the feedback channel outputs $y_{1}^{N}$ and $y_{2}^{N}$ from the previous block into three indexes $w_{0}^{*}$,
$w_{1}^{*}$ and $w_{2}^{*}$. Similar to the generalized Wyner-Ziv coding scheme introduced in Section \ref{secII},
we use the indexes $w_{0}^{*}$, $w_{1}^{*}$ and $w_{2}^{*}$ to generate $v^{N}_{1}$, $v^{N}_{2}$ and $v^{N}_{0}$, where
$v^{N}_{i}$ ($i\in\{1,2\}$) is Receiver $i$'s estimation of the channel input, and $v^{N}_{0}$ is an auxiliary sequence helping Receiver $i$ to decode
$v^{N}_{i}$. Finally, the sequence $u_{i}^{N}$ ($i\in\{1,2\}$) for each block is chosen according to the current block's $w_{i,1}$,
similar auxiliary messages $w_{i}^{'}$, $w_{i}^{''}$
shown in the non-feedback coding scheme for BC-MSR (see Section \ref{secII}), the encrypted $w_{i,2}$ and the previous block's compressed index $w_{i}^{*}$.
Moreover, the sequence $q^{N}$ is chosen according to the current block's
randomly chosen ``common message'' $w_{0}$ (see the non-feedback coding scheme for BC-MSR in Section \ref{secII})
and the previous block's compressed index $w_{0}^{*}$. Here note that for the last block, we do not transmit the real message 
$w_{i}$ ($i\in\{1,2\}$) to Receiver $i$, i.e., we transmit a constant in block $n$.

\item \emph{Decoding}: The decoding for Receiver $i$ ($i\in\{1,2\}$) begins from the last block. In block $n$, using a similar decoding scheme of 
the non-feedback coding scheme for BC-MSR, Receiver $i$ decodes $u_{i}^{N}$
and $q^{N}$ for block $n$. Then he extracts the block $n-1$'s compressed indexes $w_{i}^{*}$ and $w_{0}^{*}$ from 
the decoded $u_{i}^{N}$ and $q^{N}$ of block $n$, respectively. Next, similar to the generalized Wyner-Ziv coding scheme, Receiver $i$ views the received 
signal $y_{i}^{N}$ of block $n-1$ as side information. Given block $n-1$'s $w_{i}^{*}$, $w_{0}^{*}$ and $y_{i}^{N}$, Receiver $i$ seeks
a unique pair of $(v^{N}_{i},v^{N}_{0})$ in block $n-1$ satisfying the condition that $(v^{N}_{i},v^{N}_{0},y_{i}^{N})$ are jointly typical. Once $v^{N}_{i}$ 
of block $n-1$ is decoded, Receiver $i$ decodes $q^{N}$ for block $n-1$ by finding a unique $q^{N}$ satisfying the condition that
$(q^{N},y_{i}^{N},v^{N}_{i})$ are jointly typical. After $q^{N}$ for block $n-1$ is decoded, Receiver $i$ decodes $u_{i}^{N}$ for block $n-1$
by finding a unique $u_{i}^{N}$ satisfying the condition that
$(u_{i}^{N},q^{N},y_{i}^{N},v^{N}_{i})$ are jointly typical. Once Receiver $i$ decodes $u_{i}^{N}$
and $q^{N}$ for block $n-1$, he obtains the transmitted message $w_{i}$ for block $n-1$ and
extracts the block $n-2$'s compressed indexes $w_{i}^{*}$ and $w_{0}^{*}$.
Repeating the above decoding procedure, Receiver $i$ obtains all the messages.

\end{itemize}
Details about the proof are in Section \ref{secIV}.
\end{IEEEproof}

Then, we propose a secret key based inner bound
$\mathcal{C}^{f-in-1}_{s}$
on $\mathcal{C}^{f}_{s}$, where the feedback is used to produce keys, and these keys together with
the random binning technique
prevent each receiver's intended message from being eavesdropped by the other receiver. The inner bound $\mathcal{C}^{f-in-1}_{s}$
is shown in the following Theorem \ref{T1}.

\begin{theorem}\label{T1}
$\mathcal{C}^{f-in-1}_{s}\subseteq \mathcal{C}^{f}_{s}$,
where
\begin{eqnarray*}
&&\mathcal{C}^{f-in-1}_{s}=\{(R_{1}, R_{2}): 0\leq R_{1}\leq [I(U_{1};Y_{1}|Q)-I(U_{1};U_{2}|Q)-I(U_{1};Y_{2}|Q,U_{2})]^{+}+H(Y_{1}|Q,U_{1},U_{2},Y_{2}),\\
&&0\leq R_{2}\leq [I(U_{2};Y_{2}|Q)-I(U_{1};U_{2}|Q)-I(U_{2};Y_{1}|Q,U_{1})]^{+}+H(Y_{2}|Q,U_{1},U_{2},Y_{1}),\\
&&0\leq R_{1}\leq I(U_{1};Y_{1}|Q),\,\,0\leq R_{2}\leq I(U_{2};Y_{2}|Q),\\
&&0\leq R_{1}+R_{2}\leq \min\{I(Q;Y_{1}),I(Q;Y_{2})\}+I(U_{1};Y_{1}|Q)+I(U_{2};Y_{2}|Q)-I(U_{1};U_{2}|Q)\},
\end{eqnarray*}
and the joint distribution is denoted by
\begin{eqnarray}\label{trans2}
&&P(q,u_{1},u_{2},x,y_{1},y_{2})=P(y_{1},y_{2}|x)P(x|u_{1},u_{2})P(u_{1},u_{2}|q)P(q),
\end{eqnarray}
which indicates the Markov condition $Q\rightarrow (U_{1},U_{2})\rightarrow X\rightarrow (Y_{1},Y_{2})$.
\end{theorem}
\begin{IEEEproof}
The coding scheme achieving the inner bound $\mathcal{C}^{f-in-1}_{s}$ combines the already existing secret key based feedback scheme for the WTC \cite{AC}
with the previous non-feedback coding scheme for the BC-MSR (see Section \ref{secII}). Letting $V_{0}$, $V_{1}$ and $V_{2}$
(the estimation of the channel input) of $\mathcal{C}^{f-in-2}_{s}$ be constants, $\mathcal{C}^{f-in-1}_{s}$ is directly obtained.
Since the proof of $\mathcal{C}^{f-in-1}_{s}$ is along the lines of the proof of $\mathcal{C}^{f-in-2}_{s}$ without $V_{0}$, $V_{1}$ and $V_{2}$,
we omit the achievability proof of the inner bound $\mathcal{C}^{f-in-1}_{s}$ here.
\end{IEEEproof}

Finally, we propose an outer bound $\mathcal{C}^{f-out}_{s}$
on $\mathcal{C}^{f}_{s}$, see the following Theorem \ref{T3}.

\begin{theorem}\label{T3}
$\mathcal{C}^{f}_{s}\subseteq \mathcal{C}^{f-out}_{s}$,
where
\begin{eqnarray*}
&&\mathcal{C}^{f-out}_{s}=\{(R_{1}, R_{2}): 0\leq R_{1}\leq \min\{I(U_{1};Y_{1}|Q)-I(U_{1};Y_{2}|Q), I(U_{1};Y_{1}|Q,U_{2})-I(U_{1};Y_{2}|Q,U_{2}),\\
&&H(Y_{1}|Q,U_{2},Y_{2})\},\\
&&0\leq R_{2}\leq \min\{I(U_{2};Y_{2}|Q)-I(U_{2};Y_{1}|Q), I(U_{2};Y_{2}|Q,U_{1})-I(U_{2};Y_{1}|Q,U_{1}),
H(Y_{2}|Q,U_{1},Y_{1})\}\},
\end{eqnarray*}
the joint distribution is denoted by
\begin{eqnarray}\label{trans2.bg1}
&&P(q,u_{1},u_{2},x,y_{1},y_{2})=P(y_{1},y_{2}|x)P(x|q,u_{1},u_{2})P(q,u_{1},u_{2}),
\end{eqnarray}
and $Q$ may be assumed to be a (deterministic) function of $U_{1}$ and $U_{2}$.
\end{theorem}
\begin{IEEEproof}
See Appendix \ref{rotk1}.
\end{IEEEproof}

\section{Proof of Theorem \ref{T2}\label{secIV}}
\setcounter{equation}{0}

The messages are conveyed to the receivers via $n$ blocks.
In block $i$ ($1\leq i\leq n$), the random sequences $X^{N}$, $Y_{1}^{N}$, $Y_{2}^{N}$, $Q^{N}$, $U_{1}^{N}$, $U_{2}^{N}$, $V_{0}^{N}$, $V_{1}^{N}$
and $V_{2}^{N}$ are denoted by
$\bar{X}_{i}$, $\bar{Y}_{1,i}$, $\bar{Y}_{2,i}$, $\bar{Q}_{i}$, $\bar{U}_{1,i}$, $\bar{U}_{2,i}$, $\bar{V}_{0,i}$, $\bar{V}_{1,i}$ and $\bar{V}_{2,i}$, respectively.
In addition, let $X^{n}=(\bar{X}_{1},...,\bar{X}_{n})$ be a collection of the random sequences $X^{N}$ for all blocks.
Similarly, define $Y_{1}^{n}=(\bar{Y}_{1,1},...,\bar{Y}_{1,n})$, $Y_{2}^{n}=(\bar{Y}_{2,1},...,\bar{Y}_{2,n})$, $Q^{n}=(\bar{Q}_{1},...,\bar{Q}_{n})$,
$U_{1}^{n}=(\bar{U}_{1,1},...,\bar{U}_{1,n})$, $U_{2}^{n}=(\bar{U}_{2,1},...,\bar{U}_{2,n})$,
$V_{0}^{n}=(\bar{V}_{0,1},...,\bar{V}_{0,n})$, $V_{1}^{n}=(\bar{V}_{1,1},...,\bar{V}_{1,n})$
and $V_{2}^{n}=(\bar{V}_{2,1},...,\bar{V}_{2,n})$. The value of the random vector is written in lower case letter.

\emph{Code-books generation}:
\begin{itemize}

\item The message $W_{j}$ ($j=1,2$) is sent to Receiver $j$ via $n$ blocks, i.e., the message $W_{j}$ is composed of $n$ components
($W_{j}=(W_{j,1},...,W_{j,n})$), and
each component $W_{j,i}$ ($i\in\{1,2,...,n\}$) is the message transmitted in block $i$. Here $W_{j,i}$ takes values in
the set $\{1,...,2^{NR_{j}}\}$. Further divide $W_{j,i}$ into two parts, i.e., $W_{j,i}=(W_{j,1,i},W_{j,2,i})$. The values of $W_{j,1,i}$
and $W_{j,2,i}$ respectively belong to the sets $\{1,...,2^{NR_{j1}}\}$ and $\{1,...,2^{NR_{j2}}\}$. Here notice that $R_{j1}+R_{j2}=R_{j}$.

\item In block $i$ ($1\leq i\leq n$), randomly generate $2^{N(R_{0}+\tilde{R}_{0})}$ i.i.d. $\bar{Q}_{i}$ w.r.t. $P(q)$, and index them as
$\bar{q}_{i}(w_{0,i},w^{*}_{0,0,i},w^{*}_{0,1,i},w^{*}_{0,2,i})$, where $w_{0,i}\in \{1,2,...,2^{NR_{0}}\}$, $w^{*}_{0,0,i}\in \{1,2,...,2^{N\tilde{R}_{00}}\}$,
$w^{*}_{0,1,i}\in \{1,2,...,2^{N\tilde{R}_{01}}\}$, $w^{*}_{0,2,i}\in \{1,2,...,2^{N\tilde{R}_{02}}\}$, and
$\tilde{R}_{00}+\tilde{R}_{01}+\tilde{R}_{02}=\tilde{R}_{0}$.

\item For each possible value of $\bar{q}_{i}$, randomly generate $2^{N(R_{1}+R_{1}^{'}+R_{1}^{''}+\tilde{R}_{1})}$ i.i.d. $\bar{U}_{1,i}$ w.r.t. $P(u_{1}|q)$,
and index them as $\bar{u}_{1,i}(w_{1,1,i},w_{1,2,i},w^{'}_{1,i},w^{''}_{1,i},w^{*}_{1,0,i},w^{*}_{1,1,i})$, where
$w_{1,1,i}\in \{1,2,...,2^{NR_{11}}\}$, $w_{1,2,i}\in \{1,2,...,2^{NR_{12}}\}$, $w^{'}_{1,i}\in \{1,2,...,2^{NR^{'}_{1}}\}$,
$w^{''}_{1,i}\in \{1,2,...,2^{NR^{''}_{1}}\}$, $w^{*}_{1,0,i}\in \{1,2,...,2^{N\tilde{R}_{10}}\}$,
$w^{*}_{1,1,i}\in \{1,2,...,2^{N\tilde{R}_{11}}\}$, $R_{11}+R_{12}=R_{1}$ and $\tilde{R}_{10}+\tilde{R}_{11}=\tilde{R}_{1}$.

\item Analogously, for each possible value of $\bar{q}_{i}$, randomly generate $2^{N(R_{2}+R_{2}^{'}+R_{2}^{''}+\tilde{R}_{2})}$ i.i.d. $\bar{U}_{2,i}$ 
w.r.t. $P(u_{2}|q)$,
and index them as $\bar{u}_{2,i}(w_{2,1,i},w_{2,2,i},w^{'}_{2,i},w^{''}_{2,i},w^{*}_{2,0,i},w^{*}_{2,2,i})$, where
$w_{2,1,i}\in \{1,2,...,2^{NR_{21}}\}$, $w_{2,2,i}\in \{1,2,...,2^{NR_{22}}\}$, $w^{'}_{2,i}\in \{1,2,...,2^{NR^{'}_{2}}\}$,
$w^{''}_{2,i}\in \{1,2,...,2^{NR^{''}_{2}}\}$, $w^{*}_{2,0,i}\in \{1,2,...,2^{N\tilde{R}_{20}}\}$,
$w^{*}_{2,2,i}\in \{1,2,...,2^{N\tilde{R}_{22}}\}$, $R_{21}+R_{22}=R_{1}$ and $\tilde{R}_{20}+\tilde{R}_{22}=\tilde{R}_{2}$.

\item In block $i$, for each possible values of $\bar{q}_{i}$, $\bar{u}_{1,i}$ and $\bar{u}_{2,i}$, the channel input $\tilde{X}_{i}$ is i.i.d. generated w.r.t.
$P(x|q,u_{1},u_{2})$.

\item For each possible value of $\bar{q}_{i}$, $\bar{u}_{1,i}$, $\bar{u}_{2,i}$, $\bar{y}_{1,i}$ and $\bar{y}_{2,i}$,
produce $\bar{V}_{0,i}$ in two ways:
\begin{itemize}
\item 1)  Produce $2^{N(\tilde{R}_{00}+\tilde{R}_{0}^{'})}$ i.i.d. $\bar{V}_{0,i}$ w.r.t. 
$P(v_{0}|q,u_{1},u_{2},y_{1},y_{2})$, and index them as
$\bar{v}_{0,i}(1;w^{*}_{0,0,i},w^{*}_{1,0,i},t_{1,0,i})$, where $w^{*}_{0,0,i}\in \{1,2,...,2^{N\tilde{R}_{00}}\}$,
$w^{*}_{1,0,i}\in \{1,2,...,2^{N\tilde{R}_{10}}\}$
and $t_{1,0,i}\in \{1,2,...,2^{N(\tilde{R}_{0}^{'}-\tilde{R}_{10})}\}$.

\item Produce $2^{N(\tilde{R}_{00}+\tilde{R}_{0}^{'})}$ i.i.d. $\bar{V}_{0,i}$ w.r.t.
$P(v_{0}|q,u_{1},u_{2},y_{1},y_{2})$, and label them as
$\bar{v}_{0,i}(2;w^{*}_{0,0,i},w^{*}_{2,0,i},t_{2,0,i})$, where $w^{*}_{0,0,i}\in \{1,2,...,2^{N\tilde{R}_{00}}\}$,
$w^{*}_{2,0,i}\in \{1,2,...,2^{N\tilde{R}_{20}}\}$
and $t_{2,0,i}\in \{1,2,...,2^{N(\tilde{R}_{0}^{'}-\tilde{R}_{20})}\}$.
\end{itemize}

\item For each possible value of $\bar{q}_{i}$, $\bar{u}_{1,i}$, $\bar{u}_{2,i}$, $\bar{y}_{1,i}$ and $\bar{y}_{2,i}$, produce
$2^{N(\tilde{R}_{01}+\tilde{R}_{11}+\tilde{R}_{1}^{'})}$ i.i.d. $\bar{V}_{1,i}$ w.r.t.
$P(v_{1}|q,u_{1},u_{2},y_{1},y_{2})=\sum_{v_{0},v_{2}}P(v_{0},v_{1},v_{2}|q,u_{1},u_{2},y_{1},y_{2})$, and label them as
$\bar{v}_{1,i}(w^{*}_{0,1,i},w^{*}_{1,1,i},t_{1,i})$, where $w^{*}_{0,1,i}\in \{1,2,...,2^{N\tilde{R}_{01}}\}$,
$w^{*}_{1,1,i}\in \{1,2,...,2^{N\tilde{R}_{11}}\}$
and $t_{1,i}\in \{1,2,...,2^{N\tilde{R}_{1}^{'}}\}$.

\item Analogously, for each possible value of $\bar{q}_{i}$, $\bar{u}_{1,i}$, $\bar{u}_{2,i}$ $\bar{y}_{1,i}$ and $\bar{y}_{2,i}$, produce
$2^{N(\tilde{R}_{02}+\tilde{R}_{22}+\tilde{R}_{2}^{'})}$ i.i.d. $\bar{V}_{2,i}$ w.r.t.
$P(v_{2}|q,u_{1},u_{2},y_{1},y_{2})=\sum_{v_{0},v_{1}}P(v_{0},v_{1},v_{2}|q,u_{1},u_{2},y_{1},y_{2})$, and label them as
$\bar{v}_{2,i}(w^{*}_{0,2,i},w^{*}_{2,2,i},t_{2,i})$, where $w^{*}_{0,2,i}\in \{1,2,...,2^{N\tilde{R}_{02}}\}$,
$w^{*}_{2,2,i}\in \{1,2,...,2^{N\tilde{R}_{22}}\}$
and $t_{2,i}\in \{1,2,...,2^{N\tilde{R}_{2}^{'}}\}$.

\end{itemize}

\emph{Encoding procedure}:
\begin{itemize}

\item At block $1$, the transmitter chooses $\bar{q}_{1}(w_{0,1},1,1,1)$,
$\bar{u}_{1,1}(w_{1,1,1},w_{1,2,1}=1,w^{'}_{1,1},w^{''}_{1,1},1,1)$ and $\bar{u}_{2,1}(w_{2,1,1},w_{2,2,1}=1,w^{'}_{2,1},w^{''}_{2,1},1,1)$ to transmit.
Here notice that $w^{'}_{1,1}$ and $w^{'}_{2,1}$ are randomly chosen from the sets $\{1,2,...,2^{NR^{'}_{1}}\}$ and
$\{1,2,...,2^{NR^{'}_{2}}\}$, respectively, and the indexes $w^{''}_{1,1}$ and $w^{''}_{2,1}$ are chosen by finding a pair of
$(\bar{u}_{1,1},\bar{u}_{2,1})$ satisfying the condition that given $\bar{q}_{1}$, $(\bar{u}_{1,1},\bar{u}_{2,1},\bar{q}_{1})$ are jointly typical. If multiple pairs exist,
choose the pair with the smallest indexes; if no such pair exists, proclaim an encoding error. On the basis of the covering lemma,
this kind of encoding error tends to zero if
\begin{eqnarray}\label{sp1-r}
&&R^{''}_{1}+R^{''}_{2}\geq I(U_{1};U_{2}|Q).
\end{eqnarray}

\item At block $i$ ($i\in\{2,3,...,n-1\}$), before selecting $\bar{u}_{1,i}$ and $\bar{u}_{2,i}$, generate two mappings
$g_{1,i}: \bar{y}_{1,i-1}\rightarrow \{1,2,...,2^{NR_{12}}\}$ and $g_{2,i}: \bar{y}_{2,i-1}\rightarrow \{1,2,...,2^{NR_{22}}\}$
\footnote{Here note that these mappings are generated
exactly the same as that in \cite{AC}}.
On the basis of these two mappings, generate two RVs
$K_{1,i}=g_{1,i}(\bar{Y}_{1,i-1})$ and $K_{2,i}=g_{2,i}(\bar{Y}_{2,i-1})$ respectively taking values in $\{1,2,...,2^{NR_{12}}\}$
and $\{1,2,...,2^{NR_{22}}\}$. Here $Pr\{K_{1,i}=j\}=2^{-NR_{12}}$ for $j\in \{1,2,...,2^{NR_{12}}\}$, and
$Pr\{K_{2,i}=l\}=2^{-NR_{22}}$ for $l\in \{1,2,...,2^{NR_{22}}\}$.
The RVs $K_{1,i}$ and $K_{2,i}$ are used as secret keys encrypting the messages $w_{1,2,i}$ and $w_{2,2,i}$, respectively,
and they are independent of the transmitted messages $w_{1,2,i}$ and $w_{2,2,i}$.
The mappings $g_{1,i}$ and $g_{2,i}$ are revealed to all parties.
Once the transmitter gets $\bar{y}_{1,i-1}$ and $\bar{y}_{2,i-1}$,
he seeks a pair of $(\bar{v}_{0,i-1},\bar{v}_{1,i-1})$ satisfying the condition that
$(\bar{v}_{0,i-1}(1;\tilde{w}^{*}_{0,0,i-1},\tilde{w}^{*}_{1,0,i-1},\tilde{t}_{1,0,i-1}),\\
\bar{v}_{1,i-1}(\tilde{w}^{*}_{0,1,i-1},\tilde{w}^{*}_{1,1,i-1},\tilde{t}_{1,i-1}),\bar{u}_{1,i-1},\bar{u}_{2,i-1},\bar{q}_{i-1},\bar{y}_{1,i-1},\bar{y}_{2,i-1})$ are jointly typical.
For the case that more than one pair $(\bar{v}_{0,i-1},\bar{v}_{1,i-1})$ exist,
pick one pair at random; if there is no such pair $(\bar{v}_{0,i-1},\bar{v}_{1,i-1})$, declare an encoding error.
From the covering lemma, this kind of encoding error tends to zero if
\begin{eqnarray}\label{sp1}
&&\tilde{R}_{00}+\tilde{R}^{'}_{0}\geq I(V_{0};Q,U_{1},U_{2},Y_{1},Y_{2}),
\end{eqnarray}
\begin{eqnarray}\label{sp2}
&&\tilde{R}_{01}+\tilde{R}_{11}+\tilde{R}^{'}_{1}\geq I(V_{1};V_{0},Q,U_{1},U_{2},Y_{1},Y_{2})
\end{eqnarray}
hold. Here note that (\ref{sp1}) guarantees that there exists at least one $\bar{v}_{0,i-1}$ such that\\
$(\bar{v}_{0,i-1},\bar{u}_{1,i-1},\bar{u}_{2,i-1},\bar{q}_{i-1},\bar{y}_{1,i-1},\bar{y}_{2,i-1})$ are jointly typical, and
(\ref{sp2}) guarantees that given $\bar{v}_{0,i-1}$, there exists at least one $\bar{v}_{1,i-1}$ satisfying the condition that
$(\bar{v}_{1,i-1},\bar{v}_{0,i-1},\bar{u}_{1,i-1},\bar{u}_{2,i-1},\\ \bar{q}_{i-1},\bar{y}_{1,i-1},\bar{y}_{2,i-1})$ are jointly typical.
Similarly, the transmitter seeks a pair of $(\bar{v}_{0,i-1},\bar{v}_{2,i-1})$ satisfying the condition that
$(\bar{v}_{0,i-1}(2;\tilde{w}^{*}_{0,0,i-1},\tilde{w}^{*}_{2,0,i-1},\tilde{t}_{2,0,i-1}),
\bar{v}_{2,i-1}(\tilde{w}^{*}_{0,2,i-1},\tilde{w}^{*}_{2,2,i-1},\tilde{t}_{2,i-1}),\\ \bar{u}_{1,i-1},\bar{u}_{2,i-1},\bar{q}_{i-1},\bar{y}_{1,i-1},\bar{y}_{2,i-1})$ are jointly typical.
For the case that more than one pair $(\bar{v}_{0,i-1},\bar{v}_{2,i-1})$ exist,
pick one pair at random; if there is no such pair $(\bar{v}_{0,i-1},\bar{v}_{2,i-1})$, declare an encoding error.
From the covering lemma, this kind of encoding error tends to zero if (\ref{sp1}) and
\begin{eqnarray}\label{sp3}
&&\tilde{R}_{02}+\tilde{R}_{22}+\tilde{R}^{'}_{2}\geq I(V_{2};V_{0},Q,U_{1},U_{2},Y_{1},Y_{2})
\end{eqnarray}
hold. Once the transmitter selects the pairs $(\bar{v}_{0,i-1},\bar{v}_{1,i-1})$ and $(\bar{v}_{0,i-1},\bar{v}_{2,i-1})$, he chooses
$\bar{q}_{i}(w_{0,i},\tilde{w}^{*}_{0,0,i-1},\tilde{w}^{*}_{0,1,i-1},\tilde{w}^{*}_{0,2,i-1})$,
$\bar{u}_{1,i}(w_{1,1,i},w_{1,2,i}\oplus k_{1,i},w^{'}_{1,i},w^{''}_{1,i},\tilde{w}^{*}_{1,0,i-1},\tilde{w}^{*}_{1,1,i-1})$
and $\bar{u}_{2,i}(w_{2,1,i},w_{2,2,i}\oplus k_{2,i},w^{'}_{2,i},w^{''}_{2,i},\tilde{w}^{*}_{2,0,i-1},\tilde{w}^{*}_{2,2,i-1})$ to transmit.
Here note that $w^{'}_{1,i}$, $w^{''}_{1,i}$, $w^{'}_{2,i}$ and $w^{''}_{2,i}$ are selected the same as those in block $1$.

\item At block $n$, once the transmitter receives the feedback $\bar{y}_{1,n-1}$ and $\bar{y}_{2,n-1}$,
he seeks a pair of $(\bar{v}_{0,n-1},\bar{v}_{1,n-1})$ satisfying the condition that
$(\bar{v}_{0,n-1}(1;\tilde{w}^{*}_{0,0,n-1},\tilde{w}^{*}_{1,0,n-1},\tilde{t}_{1,0,n-1}),\\
\bar{v}_{1,n-1}(\tilde{w}^{*}_{0,1,n-1},\tilde{w}^{*}_{1,1,n-1},\tilde{t}_{1,n-1}),
\bar{u}_{1,n-1},\bar{u}_{2,n-1},\bar{q}_{n-1},\bar{y}_{1,n-1},\bar{y}_{2,n-1})$ are jointly typical, and the corresponding encoding error
tends to zero when (\ref{sp1}) and (\ref{sp2}) hold. Similarly, the transmitter seeks a pair of $(\bar{v}_{0,n-1},\bar{v}_{2,n-1})$ satisfying the condition that
$(\bar{v}_{0,n-1}(2;\tilde{w}^{*}_{0,0,n-1},\\ \tilde{w}^{*}_{2,0,n-1},\tilde{t}_{2,0,n-1}),
\bar{v}_{2,n-1}(\tilde{w}^{*}_{0,2,n-1},\tilde{w}^{*}_{2,2,n-1},\tilde{t}_{2,n-1}),
\bar{u}_{1,n-1},\bar{u}_{2,n-1},\bar{q}_{n-1},\bar{y}_{1,n-1},\bar{y}_{2,n-1})$ are jointly typical, and the corresponding encoding error
tends to zero when (\ref{sp1}) and (\ref{sp3}) hold. Then the transmitter chooses
$\bar{q}_{n}(1,\tilde{w}^{*}_{0,0,n-1},\tilde{w}^{*}_{0,1,n-1},\tilde{w}^{*}_{0,2,n-1})$,
$\bar{u}_{1,n}(1,1,1,1,\tilde{w}^{*}_{1,0,n-1},\tilde{w}^{*}_{1,1,n-1})$
and $\bar{u}_{2,n}(1,1,1,1,\tilde{w}^{*}_{2,0,n-1},\tilde{w}^{*}_{2,2,n-1})$ to transmit.

\end{itemize}

\emph{Decoding procedure}:

Receiver $j$'s ($j=1,2$) decoding procedure begins from block $n$. 
At block $n$, Receiver $j$ seeks a unique $\bar{q}_{n}$
jointly typical with $\bar{y}_{j,n}$. If there is more than one or no such $\bar{q}_{n}$, declare
an decoding error.
From the packing lemma, this kind of decoding error tends to zero if
\begin{eqnarray}\label{sp4}
&&\tilde{R}_{00}+\tilde{R}_{01}+\tilde{R}_{02}=\tilde{R}_{0}\leq I(Y_{j};Q).
\end{eqnarray}
After decoding $\bar{q}_{n}$, Receiver $j$ seeks a unique $\bar{u}_{j,n}$ satisfying the condition that $(\bar{u}_{j,n},\bar{q}_{n},\bar{y}_{j,n})$ are jointly
typical. From the packing lemma, this kind of decoding error tends to zero if
\begin{eqnarray}\label{sp5}
&&\tilde{R}_{j0}+\tilde{R}_{j1}=\tilde{R}_{j}\leq I(Y_{j};U_{j}|Q).
\end{eqnarray}
Once $\bar{q}_{n}$ and $\bar{u}_{j,n}$ are decoded, Receiver $j$ extracts $w^{*}_{0,0,n-1}$, $w^{*}_{0,1,n-1}$,
$w^{*}_{0,2,n-1}$, $w^{*}_{j,0,n-1}$ and $w^{*}_{j,j,n-1}$ in them. Then on the basis of the extracted messages,
Receiver $j$ seeks a unique pair of $(\bar{v}_{0,n-1},\bar{v}_{j,n-1})$ satisfying the condition that
$(\bar{v}_{0,n-1}(j;w^{*}_{0,0,n-1},w^{*}_{j,0,n-1},t_{j,0,n-1}),
\bar{v}_{j,n-1}(w^{*}_{0,j,n-1},\\ w^{*}_{j,j,n-1},t_{j,n-1}),\bar{y}_{j,n-1})$
are jointly typical. From the packing lemma and the multi-variate packing lemma \cite{network}, this kind of decoding error tends to zero if
\begin{eqnarray}\label{sp6}
&&\tilde{R}^{'}_{j}\leq I(V_{j};V_{0},Y_{j}),
\end{eqnarray}
\begin{eqnarray}\label{sp7}
&&\tilde{R}^{'}_{j}+\tilde{R}_{0}^{'}-\tilde{R}_{j0}\leq I(V_{0};Y_{j})+I(V_{j};V_{0},Y_{j}).
\end{eqnarray}
Here notice that (\ref{sp6}) guarantees the probability of the error event that there is a unique $\bar{v}_{0,n-1}$ and more than one $t_{j,n-1}$ satisfying the condition that
$(\bar{v}_{0,n-1},\bar{v}_{j,n-1},\bar{y}_{j,n-1})$ are jointly typical tends to zero, and (\ref{sp7}) guarantees
the probability of the error event that there exist more than one $t_{j,0,n-1}$ and $t_{j,n-1}$
satisfying the condition that $(\bar{v}_{0,n-1},\bar{v}_{j,n-1},\bar{y}_{j,n-1})$ are jointly typical tends to zero.
Then, for block $n-1$, after $\bar{v}_{0,n-1}$ and $\bar{v}_{j,n-1}$ are decoded, Receiver $j$ seeks a unique $\bar{q}_{n-1}$ satisfying the condition that
$(\bar{q}_{n-1},\bar{y}_{j,n-1},\bar{v}_{j,n-1})$ are jointly typical, and from the packing lemma \cite{network},
this kind of decoding error tends to zero if
\begin{eqnarray}\label{sp8}
&&R_{0}+\tilde{R}_{0}\leq I(Q;V_{j},Y_{j}).
\end{eqnarray}
Then Receiver $j$ seeks a unique $\bar{u}_{j,n-1}$ satisfying the condition that $(\bar{u}_{j,n-1},\bar{q}_{n-1},\bar{y}_{j,n-1},\bar{v}_{j,n-1})$ 
are jointly typical,
and this kind of decoding error tends to zero if
\begin{eqnarray}\label{sp9}
&&R_{j1}+R_{j2}+R_{j}^{'}+R_{j}^{''}+\tilde{R}_{j0}+\tilde{R}_{j,j}\leq I(U_{j};V_{j},Y_{j}|Q).
\end{eqnarray}
When $\bar{q}_{n-1}$ and $\bar{u}_{j,n-1}$ are decoded, Receiver $j$ extracts $w_{0,n-1}$, $w^{*}_{0,0,n-2}$
$w^{*}_{0,1,n-2}$, $w^{*}_{0,2,n-2}$, $w_{j,1,n-1}$, $w_{j,2,n-1}\oplus k_{j,n-1}$, $w^{'}_{j,n-1}$, $w^{''}_{j,n-1}$,
$w^{*}_{j,0,n-2}$ and $w^{*}_{j,j,n-2}$. Since the key $k_{j,n-1}$ generated from $\bar{y}_{j,n-2}$ is also known by Receiver $j$,
the messages $w_{j,1,n-1}$ and $w_{j,2,n-1}$ are correctly decoded by Receiver $j$. Repeat the above decoding procedure, the messages of all blocks are
decoded by Receiver $j$.
The following Figs.\ref{t1}-\ref{t3} help us to better understand the proposed encoding and decoding schemes¡£

\begin{figure}[htb]
\centering
\includegraphics[scale=0.4]{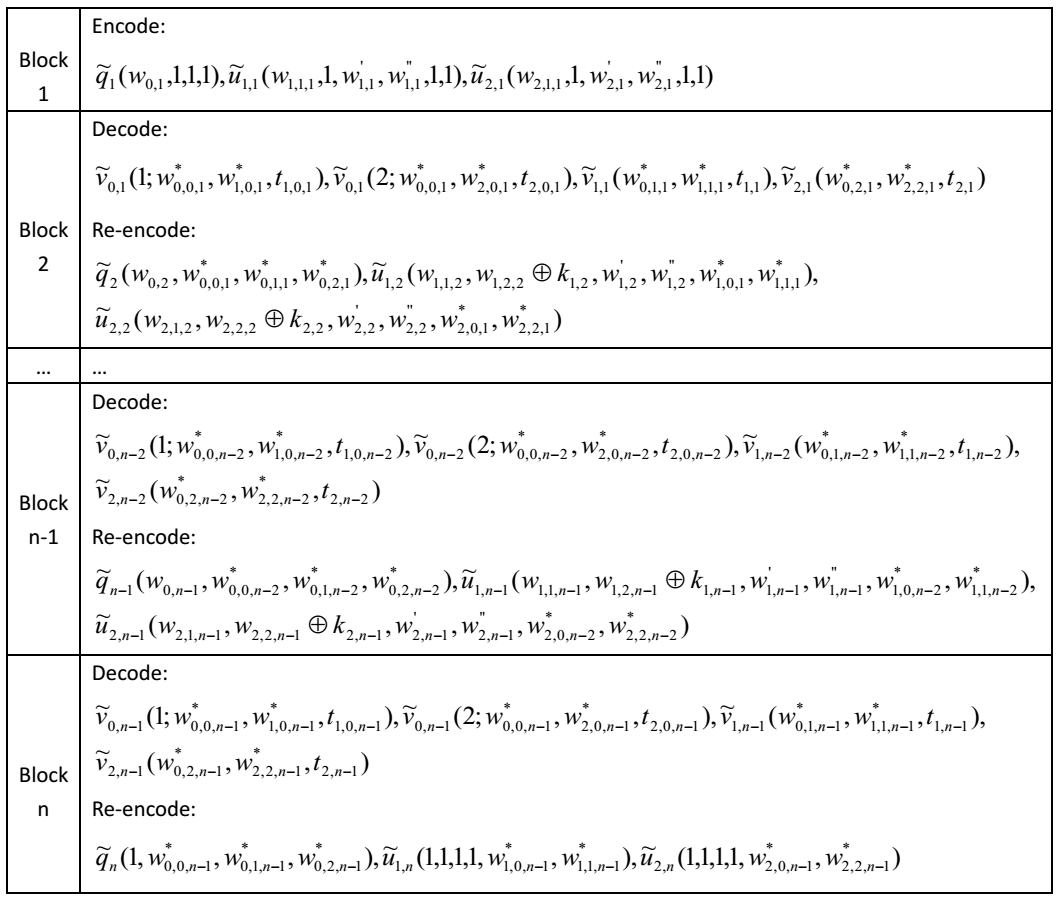}
\caption{The encoding procedure of the transmitter}
\label{t1}
\end{figure}

\begin{figure}[htb]
\centering
\includegraphics[scale=0.4]{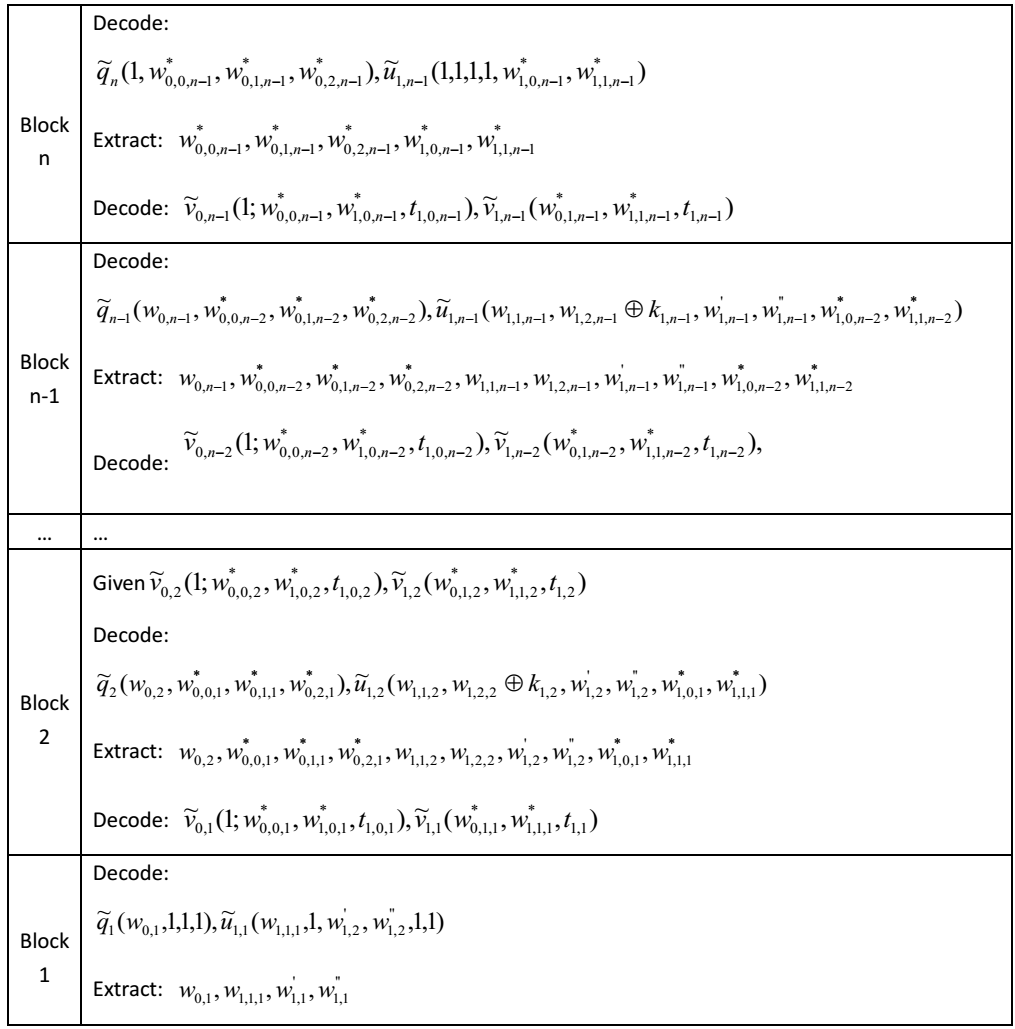}
\caption{The decoding procedure of Receiver $1$}
\label{t2}
\end{figure}

\begin{figure}[htb]
\centering
\includegraphics[scale=0.4]{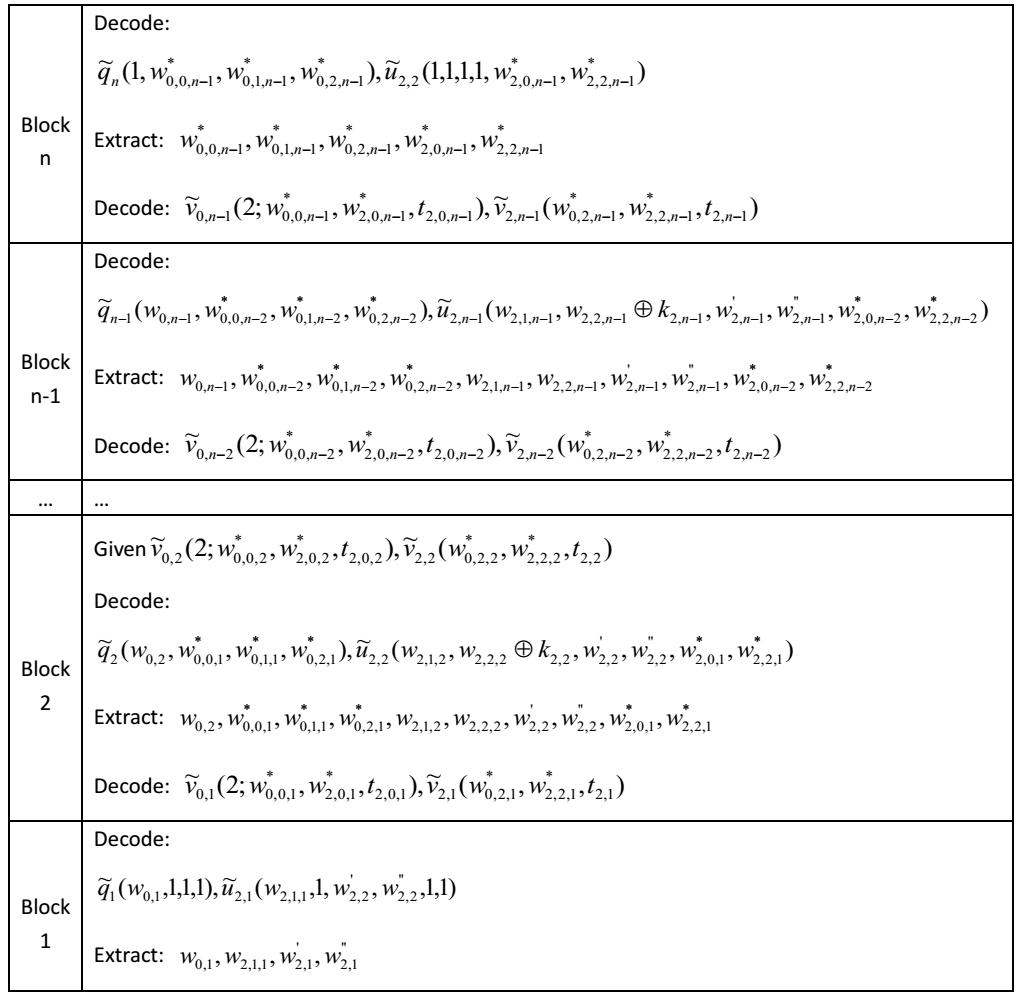}
\caption{The decoding procedure of Receiver $2$}
\label{t3}
\end{figure}

\emph{Equivocation Analysis for Receiver $2$}:

Receiver $2$'s equivocation $\Delta_{1}$, which is denoted by $\Delta_{1}=\frac{1}{nN}H(W_{1}|Y_{2}^{n})$, follows that
\begin{eqnarray}\label{jsac1}
&&\Delta_{1}=\frac{1}{nN}H(W_{1}|Y_{2}^{n})\stackrel{(a)}=\frac{1}{nN}H(\tilde{W}_{11},\tilde{W}_{12}|Y_{2}^{n})\nonumber\\
&&=\frac{1}{nN}(H(\tilde{W}_{11}|Y_{2}^{n})+H(\tilde{W}_{12}|Y_{2}^{n},\tilde{W}_{11})),
\end{eqnarray}
where (a) follows from the definitions $\tilde{W}_{11}=(W_{1,1,1},...,W_{1,1,n})$ and $\tilde{W}_{12}=(W_{1,2,1},...,W_{1,2,n})$.

The first term $H(\tilde{W}_{11}|Y_{2}^{n})$ of (\ref{jsac1}) follows that
\begin{eqnarray}\label{jsac2}
&&H(\tilde{W}_{11}|Y_{2}^{n})\geq H(\tilde{W}_{11}|Y_{2}^{n},Q^{n},U_{2}^{n},V_{2}^{n})\nonumber\\
&&=H(\tilde{W}_{11},Y_{2}^{n},Q^{n},U_{2}^{n},V_{2}^{n})-H(Y_{2}^{n},Q^{n},U_{2}^{n},V_{2}^{n})\nonumber\\
&&=H(\tilde{W}_{11},Y_{2}^{n},Q^{n},U_{2}^{n},V_{2}^{n},U_{1}^{n})-H(U_{1}^{n}|\tilde{W}_{11},Y_{2}^{n},Q^{n},U_{2}^{n},V_{2}^{n})
-H(Y_{2}^{n},Q^{n},U_{2}^{n},V_{2}^{n})\nonumber\\
&&\stackrel{(b)}=H(Y_{2}^{n},V_{2}^{n}|Q^{n},U_{2}^{n},U_{1}^{n})+H(Q^{n},U_{2}^{n},U_{1}^{n})-H(U_{1}^{n}|\tilde{W}_{11},Y_{2}^{n},Q^{n},U_{2}^{n},V_{2}^{n})\nonumber\\
&&-H(Y_{2}^{n},V_{2}^{n}|Q^{n},U_{2}^{n})-H(Q^{n},U_{2}^{n})\nonumber\\
&&=H(Y_{2}^{n},V_{2}^{n}|Q^{n},U_{2}^{n},U_{1}^{n})+H(U_{1}^{n}|Q^{n},U_{2}^{n})-H(U_{1}^{n}|\tilde{W}_{11},Y_{2}^{n},Q^{n},U_{2}^{n},V_{2}^{n})\nonumber\\
&&-H(Y_{2}^{n},V_{2}^{n}|Q^{n},U_{2}^{n})\nonumber\\
&&=H(U_{1}^{n}|Q^{n})-I(U_{1}^{n};U_{2}^{n}|Q^{n})-H(U_{1}^{n}|\tilde{W}_{11},Y_{2}^{n},Q^{n},U_{2}^{n},V_{2}^{n})\nonumber\\
&&-I(U_{1}^{n};Y_{2}^{n},V_{2}^{n}|Q^{n},U_{2}^{n})\nonumber\\
&&\stackrel{(c)}=H(U_{1}^{n}|Q^{n})-nNI(U_{1};U_{2}|Q)-nNI(U_{1};Y_{2},V_{2}|Q,U_{2})-H(U_{1}^{n}|\tilde{W}_{11},Y_{2}^{n},Q^{n},U_{2}^{n},V_{2}^{n})\nonumber\\
&&\stackrel{(d)}=(n-1)N(R_{11}+R_{1}^{'}+R_{1}^{''}+\tilde{R}_{10}+\tilde{R}_{11})+(n-2)NR_{12}-nNI(U_{1};U_{2}|Q)\nonumber\\
&&-nNI(U_{1};Y_{2},V_{2}|Q,U_{2})-H(U_{1}^{n}|\tilde{W}_{11},Y_{2}^{n},Q^{n},U_{2}^{n},V_{2}^{n})\nonumber\\
&&\stackrel{(e)}\geq (n-1)N(R_{11}+R_{1}^{'}+R_{1}^{''}+\tilde{R}_{10}+\tilde{R}_{11})+(n-2)NR_{12}\nonumber\\
&&-nNI(U_{1};U_{2}|Q)-nNI(U_{1};Y_{2},V_{2}|Q,U_{2})-nN\epsilon_{3},
\end{eqnarray}
where (b) follows from $H(\tilde{W}_{11}|U_{1}^{n})=0$, (c) follows from the generation of $Q^{n}$, $U_{1}^{n}$, $U_{2}^{n}$, $V_{1}^{n}$, $V_{2}^{n}$ and
the channel is memoryless,
and (d) follows from the generation of $U_{1}^{n}$, and (e) follows from
that given $\tilde{w}_{11}$, $y_{2}^{n}$, $v_{2}^{n}$, $q^{n}$ and $u_{2}^{n}$, Receiver $2$ seeks a unique 
$u_{1}^{n}$ that is jointly typical with his own received
signals $y_{2}^{n}$, $v_{2}^{n}$, $q^{n}$, $u_{2}^{n}$, and
from the packing lemma, we see that
Receiver $2$'s decoding error tends to zero if
\begin{eqnarray}\label{jsac3}
&&R_{12}+R_{1}^{'}+\tilde{R}_{10}+\tilde{R}_{11}\leq I(Y_{2},V_{2};U_{1}|Q,U_{2}),
\end{eqnarray}
then applying Fano's lemma, $\frac{1}{nN}H(U_{1}^{n}|\tilde{W}_{11},Y_{2}^{n},Q^{n},U_{2}^{n},V_{2}^{n})\leq \epsilon_{3}$ is obtained,
where $\epsilon_{3}\rightarrow 0$ while $n, N\rightarrow \infty$.

The second term $H(\tilde{W}_{12}|Y_{2}^{n},\tilde{W}_{11})$ of (\ref{jsac1}) is bounded by
\begin{eqnarray}\label{jsac4}
&&H(\tilde{W}_{12}|Y_{2}^{n},\tilde{W}_{11})\nonumber\\
&&\geq \sum_{i=2}^{n-1}H(W_{1,2,i}|Y_{2}^{n},\tilde{W}_{11},W_{1,2,1}=1,...,W_{1,2,i-1},W_{1,2,i}\oplus K_{1,i})\nonumber\\
&&\stackrel{(f)}=\sum_{i=2}^{n-1}H(W_{1,2,i}|\bar{Y}_{2,i-1},W_{1,2,i}\oplus K_{1,i})\nonumber\\
&&\geq \sum_{i=2}^{n-1}H(W_{1,2,i}|\bar{Y}_{2,i-1},W_{1,2,i}\oplus K_{1,i},\bar{V}_{2,i-1},\bar{Q}_{i-1},\bar{U}_{1,i-1},\bar{U}_{2,i-1})\nonumber\\
&&=\sum_{i=2}^{n-1}H(K_{1,i}|\bar{Y}_{2,i-1},W_{1,2,i}\oplus K_{1,i},\bar{V}_{2,i-1},\bar{Q}_{i-1},\bar{U}_{1,i-1},\bar{U}_{2,i-1})\nonumber\\
&&\stackrel{(g)}=\sum_{i=2}^{n-1}H(K_{1,i}|\bar{Y}_{2,i-1},\bar{V}_{2,i-1},\bar{Q}_{i-1},\bar{U}_{1,i-1},\bar{U}_{2,i-1})\nonumber\\
&&\stackrel{(h)}\geq (n-2)(\log\frac{1-\epsilon_{1}}{1+\delta}+N(1-\epsilon_{2})H(Y_{1}|Y_{2},V_{2},Q,U_{1},U_{2})),
\end{eqnarray}
where (f) follows from the Markov chain $W_{1,2,i}\rightarrow (\bar{Y}_{2,i-1},W_{1,2,i}\oplus K_{1,i})
\rightarrow (\tilde{W}_{11},W_{1,2,1},...,W_{1,2,i-1},\\\bar{Y}_{2,1},...,\bar{Y}_{2,i-2},
\bar{Y}_{2,i},...,\bar{Y}_{2,n})$, (g) follows from
$K_{1,i}\rightarrow (\bar{Y}_{2,i-1},\bar{V}_{2,i-1},\bar{Q}_{i-1},\bar{U}_{1,i-1},\bar{U}_{2,i-1})\rightarrow W_{1,2,i}\oplus K_{1,i}$,
and (h) follows from Lemma \ref{L4} that given $\bar{y}_{2,i-1}$, $\bar{q}_{i-1}$, $\bar{u}_{1,i-1}$, $\bar{u}_{2,i-1}$ and $\bar{v}_{2,i-1}$, there are at least
$\frac{\gamma}{1+\delta}$ colors (see (\ref{aaa2})), which indicates that
\begin{eqnarray}\label{zhenni2}
&&H(K_{1,i}|\bar{Y}_{2,i-1},\bar{V}_{2,i-1},\bar{Q}_{i-1},\bar{U}_{1,i-1},\bar{U}_{2,i-1})
\geq \log\frac{\gamma}{1+\delta},
\end{eqnarray}
then substituting (\ref{zhenni1}) into (\ref{zhenni2}), we get
\begin{eqnarray}\label{zhenni3}
&&H(K_{1,i}|\bar{Y}_{2,i-1},\bar{V}_{2,i-1},\bar{Q}_{i-1},\bar{U}_{1,i-1},\bar{U}_{2,i-1})\nonumber\\
&&\geq \log\frac{1-\epsilon_{1}}{1+\delta}+N(1-\epsilon_{2})H(Y_{1}|Y_{2},V_{2},Q,U_{1},U_{2}),
\end{eqnarray}
where $\epsilon_{1}$, $\epsilon_{2}$ and $\delta$ tend to $0$ while $N$ goes to infinity.

Substituting (\ref{jsac4}) and (\ref{jsac2}) into (\ref{jsac1}), we get
\begin{eqnarray}\label{jsac5}
&&\Delta_{1}\geq \frac{n-1}{n}(R_{11}+R_{1}^{'}+R_{1}^{''}+\tilde{R}_{10}+\tilde{R}_{11})+\frac{n-2}{n}R_{12}-I(U_{1};U_{2}|Q)\nonumber\\
&&-I(V_{2},Y_{2};U_{1}|Q,U_{2})-\epsilon_{3}+\frac{n-2}{nN}\log\frac{1-\epsilon_{1}}{1+\delta}+\frac{n-2}{n}(1-\epsilon_{2})H(Y_{1}|Y_{2},V_{2},Q,U_{1},U_{2}).\nonumber\\
\end{eqnarray}
The bound (\ref{jsac5}) implies that if
\begin{eqnarray}\label{jsac6}
&&R_{1}^{'}+R_{1}^{''}+\tilde{R}_{10}+\tilde{R}_{11}\geq I(U_{1};U_{2}|Q)+I(V_{2},Y_{2};U_{1}|Q,U_{2})-H(Y_{1}|Y_{2},V_{2},Q,U_{1},U_{2}),\nonumber\\
\end{eqnarray}
$\Delta_{1}\geq R_{1}-\epsilon$ is obtained
by choosing sufficiently large $n$ and $N$.

\emph{Equivocation Analysis for Receiver $1$}:
The equivocation analysis of Receiver $1$'s equivocation $\Delta_{2}=\frac{1}{nN}H(W_{2}|Y_{1}^{n})$ is analogous to that of $\Delta_{1}$, hence
$\Delta_{2}\geq R_{2}-\epsilon$ can be proved by letting
\begin{eqnarray}\label{jsac3-1}
&&R_{22}+R_{2}^{'}+\tilde{R}_{20}+\tilde{R}_{22}\leq I(Y_{1},V_{1};U_{2}|Q,U_{1}),
\end{eqnarray}
\begin{eqnarray}\label{jsac6-1}
&&R_{2}^{'}+R_{2}^{''}+\tilde{R}_{20}+\tilde{R}_{22}\nonumber\\
&&\geq I(U_{1};U_{2}|Q)+I(V_{1},Y_{1};U_{2}|Q,U_{1})-H(Y_{2}|Y_{1},V_{1},Q,U_{1},U_{2}),
\end{eqnarray}
and selecting sufficiently large $n$ and $N$.

Now using the fact that $\tilde{R}_{0}=\tilde{R}_{00}+\tilde{R}_{01}+\tilde{R}_{02}$, $\tilde{R}_{1}=\tilde{R}_{10}+\tilde{R}_{11}$,
$\tilde{R}_{2}=\tilde{R}_{20}+\tilde{R}_{22}$, and applying Fourier-Motzkin elimination to remove $\tilde{R}_{0}^{'}$, $\tilde{R}_{1}^{'}$ and $\tilde{R}_{2}^{'}$
from (\ref{sp1})-(\ref{sp7}), we have
\begin{eqnarray}\label{malong1}
&&\tilde{R}_{0}+\tilde{R}_{1}\geq I(V_{0},V_{1};Q,U_{1},U_{2},Y_{1},Y_{2}|Y_{1})\label{malong1-1}\\
&&\tilde{R}_{0}+\tilde{R}_{2}\geq I(V_{0},V_{2};Q,U_{1},U_{2},Y_{1},Y_{2}|Y_{2})\label{malong1-2}\\
&&\tilde{R}_{0}+\tilde{R}_{1}+\tilde{R}_{2}\geq I(V_{1};Q,U_{1},U_{2},Y_{1},Y_{2}|Y_{1},V_{0})+I(V_{2};Q,U_{1},U_{2},Y_{1},Y_{2}|Y_{2},V_{0})\nonumber\\
&&+\max\{I(V_{0};Q,U_{1},U_{2},Y_{1},Y_{2}|Y_{1}),I(V_{0};Q,U_{1},U_{2},Y_{1},Y_{2}|Y_{2})\}.\label{malong1-3}
\end{eqnarray}
Then, further using the fact that 
$R_{1}=R_{11}+R_{12}$ and $R_{2}=R_{21}+R_{22}$, and applying
Fourier-Motzkin elimination to remove $R_{1}^{'}$, $R_{1}^{''}$, $R_{2}^{'}$, $R_{2}^{''}$, $\tilde{R}_{0}$, $\tilde{R}_{1}$ and $\tilde{R}_{2}$
from (\ref{malong1-1}), (\ref{malong1-2}), (\ref{malong1-3}), (\ref{sp1-r}),
(\ref{sp8}), (\ref{sp9}), (\ref{jsac3}), (\ref{jsac6}), (\ref{jsac3-1}) and (\ref{jsac6-1}),
Theorem \ref{T2} is proved.

\section{Examples}\label{secV}
\setcounter{equation}{0}

\subsection{Dueck-type Example}\label{3-1}

\begin{figure}[htb]
\centering
\includegraphics[scale=0.4]{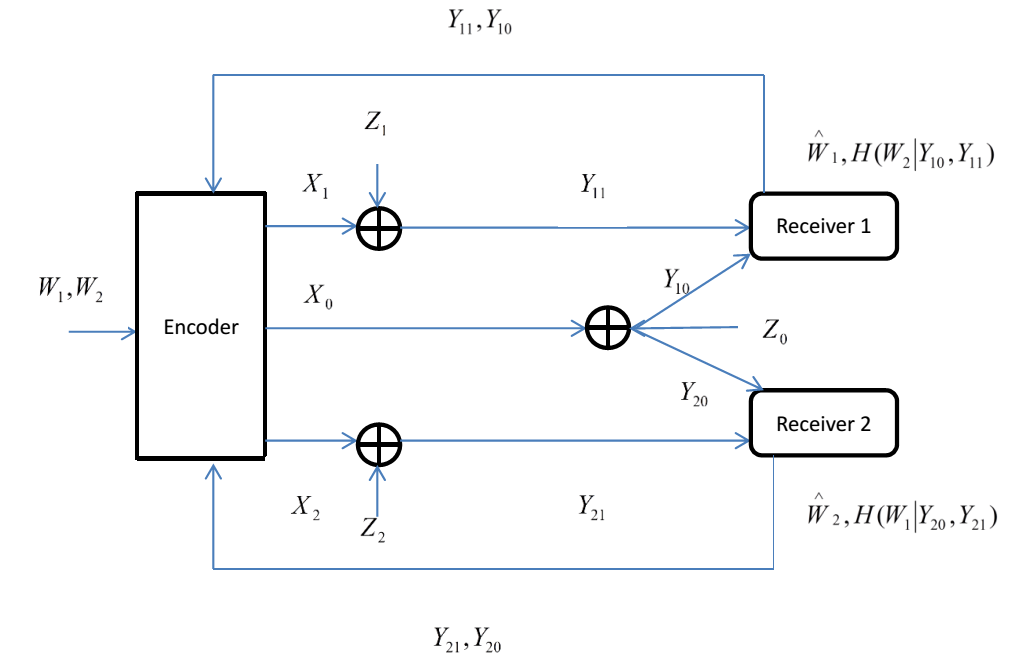}
\caption{A Dueck-type example of the BC-MSR with noiseless feedback}
\label{f2}
\end{figure}

In this subsection, we further explain the inner and outer bounds on the secrecy capacity region of the BC-MSR with noiseless feedback via a Dueck-type example.
In this example (see Figure \ref{f2}), the channel input and outputs satisfy
\begin{eqnarray}\label{s3}
&&X=(X_{0},X_{1},X_{2}),\,Y_{1}=(Y_{10},Y_{11}),\,Y_{2}=(Y_{20},Y_{21}),\nonumber\\
&&Y_{10}=Y_{20}=X_{0}\oplus Z_{0},\,Y_{11}=X_{1}\oplus Z_{1},\,Y_{21}=X_{2}\oplus Z_{2},
\end{eqnarray}
where the channel inputs $X_{0}$, $X_{1}$, $X_{2}$ and the channel noises $Z_{0}$, $Z_{1}$, $Z_{2}$ are binary random variables (taking values in $\{0,1\}$),
and the channel noises are independent of the channel inputs.

First, we show the secret key based inner bound $\mathcal{C}^{f-in-1}_{s}$ on the secrecy capacity region of this Dueck-type example.
Letting
\begin{eqnarray}\label{s4.1}
&&P(X_{0}=0)=\alpha_{1},\,P(X_{0}=1)=1-\alpha_{1},\nonumber\\
&&P(X_{1}=0)=\alpha_{2},\,P(X_{1}=1)=1-\alpha_{2},\nonumber\\
&&P(X_{2}=0)=\alpha_{3},\,P(X_{2}=1)=1-\alpha_{3},
\end{eqnarray}
and substituting $Q=X_{0}$, $U_{1}=X_{1}$,
$U_{2}=X_{2}$ and (\ref{s3}) into Theorem \ref{T1}, $\mathcal{C}^{f-in-1}_{s}$ reduces to $\mathcal{C}^{in-1}_{sf}$, and it is given by
\begin{eqnarray}\label{s4}
&&\mathcal{C}^{in-1}_{sf}=\{(R_{1}, R_{2}): R_{1}\leq 1-H(Z_{1}|Z_{0})+H(Z_{1}|Z_{0},Z_{2}),\nonumber\\
&&R_{2}\leq 1-H(Z_{2}|Z_{0})+H(Z_{2}|Z_{0},Z_{1}),\nonumber\\
&&R_{1}\leq 2-H(Z_{0},Z_{1}),\, R_{2}\leq 2-H(Z_{0},Z_{2}),\nonumber\\
&&R_{1}+R_{2}\leq 3+H(Z_{0})-H(Z_{0},Z_{1})-H(Z_{0},Z_{2})\}.
\end{eqnarray}
Here note that (\ref{s4}) is obtained when $\alpha_{1}=\alpha_{2}=\alpha_{3}=\frac{1}{2}$.

Second, we show the hybrid inner bound $\mathcal{C}^{f-in-2}_{s}$ on the secrecy capacity region of this Dueck-type example. Substituting (\ref{s4.1}), $Q=X_{0}$, $U_{1}=X_{1}$,
$U_{2}=X_{2}$, $V_{1}=(X_{0},X_{1})$, $V_{2}=(X_{0},X_{2})$, (\ref{s3}), $V_{0}=(Z_{0},Z_{1})$ or $V_{0}=(Z_{0},Z_{2})$ into
Theorem \ref{T2}, $\mathcal{C}^{f-in-2}_{s}$ reduces to $\mathcal{C}^{in-2}_{sf}$, and it is given by
\begin{eqnarray}\label{s5}
&&\mathcal{C}^{in-2}_{sf}=\{(R_{1}, R_{2}): R_{1}\leq 1+H(Z_{1}|Z_{0},Z_{2}),\,R_{2}\leq 1+H(Z_{2}|Z_{0},Z_{1},\nonumber\\
&&R_{1}\leq 2-H(Z_{0},Z_{1}),\,R_{2}\leq 2-H(Z_{0},Z_{2}),\nonumber\\
&&R_{1}+R_{2}\leq 3-H(Z_{0},Z_{1},Z_{2})\}.
\end{eqnarray}

Third, we show a simple cut-set outer bound $\mathcal{C}^{out}_{sf}$ on the secrecy capacity region of this Dueck-type example.
Since only $X_{0}$ and $X_{1}$ are transmitted to receiver $1$, the transmission rate $R_{1}$ of the message $W_{1}$
is upper bounded by $I(X_{0},X_{1};Y_{1})$. Analogously, the transmission rate $R_{2}$
is upper bounded by $I(X_{0},X_{2};Y_{2})$. For all receivers, the sum rate $R_{1}+R_{2}$ is upper bounded by
$I(X_{0},X_{1},X_{2};Y_{1},Y_{2})$. Now it remains to calculate these upper bounds.
Since $X_{0}$, $X_{1}$, $X_{2}$, $Z_{0}$, $Z_{1}$ and $Z_{2}$ take values in $\{0,1\}$, from (\ref{s3}), we have
\begin{eqnarray}\label{s6}
&&I(X_{0},X_{1};Y_{1})=H(X_{0}\oplus Z_{0},X_{1}\oplus Z_{1})-H(X_{0}\oplus Z_{0},X_{1}\oplus Z_{1}|X_{0},X_{1})\nonumber\\
&&=H(X_{0}\oplus Z_{0},X_{1}\oplus Z_{1})-H(Z_{0},Z_{1}|X_{0},X_{1})\nonumber\\
&&\stackrel{(1)}=H(X_{0}\oplus Z_{0},X_{1}\oplus Z_{1})-H(Z_{0},Z_{1})\nonumber\\
&&\leq H(X_{0}\oplus Z_{0})+H(X_{1}\oplus Z_{1})-H(Z_{0},Z_{1})\nonumber\\
&&\leq 2-H(Z_{0},Z_{1}),
\end{eqnarray}
where (1) follows from the fact that the channel noises are independent of the channel inputs.
Analogously, we have
\begin{eqnarray}\label{s7}
&&I(X_{0},X_{2};Y_{2})\leq 2-H(Z_{0},Z_{2}).
\end{eqnarray}
For $I(X_{0},X_{1},X_{2};Y_{1},Y_{2})$, we have
\begin{eqnarray}\label{s8}
&&I(X_{0},X_{1},X_{2};Y_{1},Y_{2})\nonumber\\
&&=H(X_{0}\oplus Z_{0},X_{1}\oplus Z_{1},X_{2}\oplus Z_{2})
-H(X_{0}\oplus Z_{0},X_{1}\oplus Z_{1},X_{2}\oplus Z_{2}|X_{0},X_{1},X_{2})\nonumber\\
&&=H(X_{0}\oplus Z_{0},X_{1}\oplus Z_{1},X_{2}\oplus Z_{2})-H(Z_{0},Z_{1},Z_{2}|X_{0},X_{1},X_{2})\nonumber\\
&&\stackrel{(2)}=H(X_{0}\oplus Z_{0},X_{1}\oplus Z_{1},X_{2}\oplus Z_{2})-H(Z_{0},Z_{1},Z_{2})\nonumber\\
&&\leq H(X_{0}\oplus Z_{0})+H(X_{1}\oplus Z_{1})+H(X_{2}\oplus Z_{2})-H(Z_{0},Z_{1},Z_{2})\nonumber\\
&&\leq 3-H(Z_{0},Z_{1},Z_{2}),
\end{eqnarray}
where (2) also follows from the fact that the channel noises are independent of the channel inputs.
Combining (\ref{s6}), (\ref{s7}) with (\ref{s8}),
an outer bound $\mathcal{C}^{out}_{sf}$ on the secrecy capacity region
of this Dueck-type example is given by
\begin{eqnarray}\label{s9}
&&\mathcal{C}^{out}_{sf}=\{(R_{1}, R_{2}): R_{1}\leq 2-H(Z_{0},Z_{1}),\,R_{2}\leq 2-H(Z_{0},Z_{2}),\nonumber\\
&&R_{1}+R_{2}\leq 3-H(Z_{0},Z_{1},Z_{2})\}.
\end{eqnarray}

Finally, in order to show the advantage of feedback, we also give an inner bound $\mathcal{C}^{in}_{s}$ on the secrecy capacity region
of the Dueck-type example without the feedback. Here notice that in \cite{bc1}, an achievable secrecy rate region $\mathcal{C}^{in*}_{s}$
for the BC-MSR is proposed, and it is given by
\begin{eqnarray}\label{s10}
&&\mathcal{C}^{in*}_{s}=\{(R_{1}, R_{2}): 0\leq R_{1}\leq I(U_{1};Y_{1}|Q)-I(U_{1};U_{2}|Q)-I(U_{1};Y_{2}|Q,U_{2}),\nonumber\\
&&0\leq R_{2}\leq I(U_{2};Y_{2}|Q)-I(U_{1};U_{2}|Q)-I(U_{2};Y_{1}|Q,U_{1})\}.
\end{eqnarray}
Now substituting $Q=X_{0}$, $U_{1}=X_{1}$,
$U_{2}=X_{2}$, (\ref{s4.1}) and (\ref{s3}) into (\ref{s10}),
$\mathcal{C}^{in*}_{s}$ reduces to $\mathcal{C}^{in}_{s}$, and it is given by
\begin{eqnarray}\label{s11}
&&\mathcal{C}^{in}_{s}=\{(R_{1}, R_{2}): R_{1}\leq 1-H(Z_{1}|Z_{0}),\,R_{2}\leq 1-H(Z_{2}|Z_{0})\}.
\end{eqnarray}
Here note that (\ref{s11}) is obtained when $\alpha_{1}=\alpha_{2}=\alpha_{3}=\frac{1}{2}$.

In order to compare the above bounds, we further define the channel noises $Z_{0}$, $Z_{1}$ and $Z_{2}$ in the following two cases:
\begin{itemize}
\item Case 1: The channel noise $Z_{2}$ only depends on $Z_{1}$, i.e., there exists a Markov chain
$Z_{0}\rightarrow Z_{1}\rightarrow Z_{2}$.

\item Case 2: The channel noises $Z_{1}$ and $Z_{2}$ only depend on the noise $Z_{0}$, i.e., there exists a Markov chain
$Z_{1}\rightarrow Z_{0}\rightarrow Z_{2}$.
\end{itemize}

In the remainder of this subsection, we show the numerical results on the above two cases of this Dueck-type example, see the followings.

\subsubsection{A special case of Dueck-type example with $Z_{0}\rightarrow Z_{1}\rightarrow Z_{2}$}\label{3-1-1}

\begin{figure}[htb]
\centering
\includegraphics[scale=0.4]{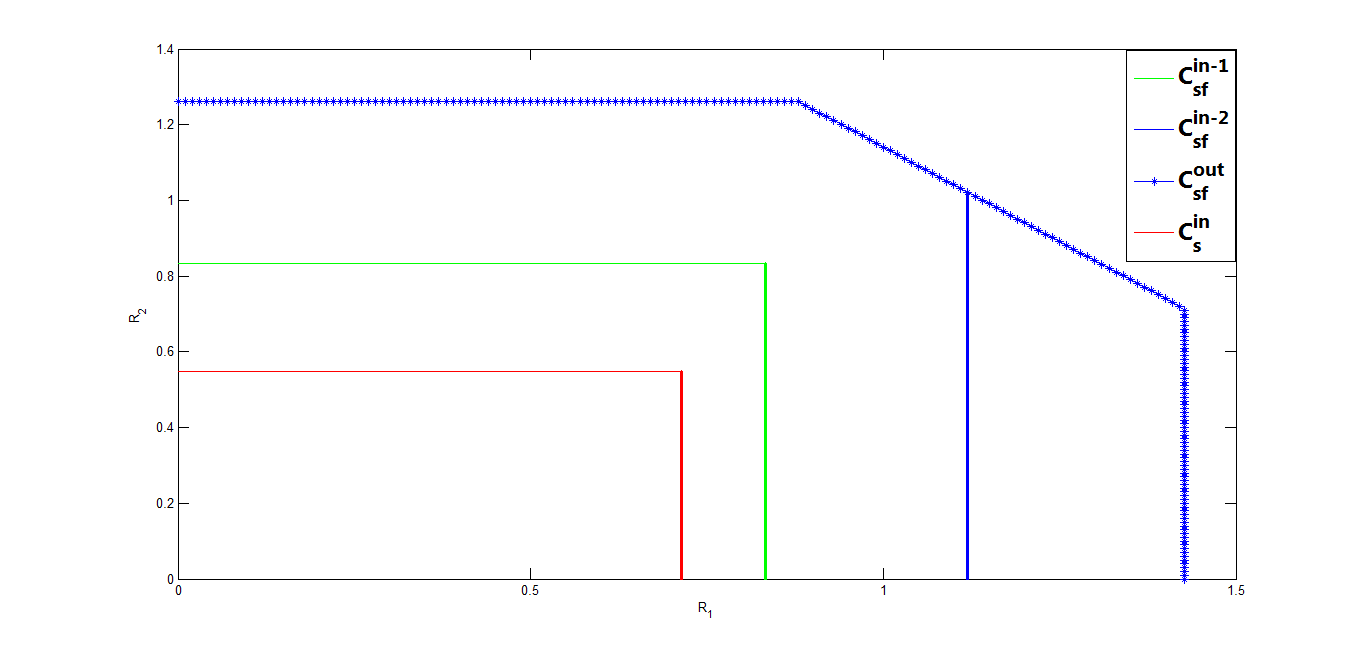}
\caption{Comparison of the bounds for the case $Z_{0}\rightarrow Z_{1}\rightarrow Z_{2}$ and $p=q=r=0.05$}
\label{fr1}
\end{figure}

For the case $Z_{0}\rightarrow Z_{1}\rightarrow Z_{2}$, define
\begin{eqnarray}\label{s12}
\left\{
\begin{array}{ll}
P(Z_{0}=0)=1-p,\,P(Z_{0}=1)=p,\\
P(Z_{1}=0|Z_{0}=0)=1-q,\,P(Z_{1}=1|Z_{0}=0)=q,\\
P(Z_{1}=0|Z_{0}=1)=q,\,P(Z_{1}=1|Z_{0}=1)=1-q,\\
P(Z_{2}=0|Z_{1}=0)=1-r,\,P(Z_{2}=1|Z_{1}=0)=r,\\
P(Z_{2}=0|Z_{1}=1)=r,\,P(Z_{2}=1|Z_{1}=1)=1-r
\end{array}
\right\},
\end{eqnarray}
where $0\leq p,q,r\leq \frac{1}{2}$. Substituting (\ref{s12}) and $Z_{0}\rightarrow Z_{1}\rightarrow Z_{2}$
into (\ref{s4}), (\ref{s5}), (\ref{s9}) and (\ref{s11}),
we have
\begin{eqnarray}\label{s4-r}
&&\mathcal{C}^{in-1}_{sf}=\{(R_{1}, R_{2}): R_{1}\leq 1-h(r\star q)+h(r),\,R_{2}\leq 1-h(r\star q)+h(r),\nonumber\\
&&R_{1}\leq 2-h(p)-h(q),\,R_{2}\leq 2-h(p)-h(r\star q),\nonumber\\
&&R_{1}+R_{2}\leq 3-h(q)-h(p)-h(r\star q)\},
\end{eqnarray}
\begin{eqnarray}\label{s5-r}
&&\mathcal{C}^{in-2}_{sf}=\{(R_{1}, R_{2}): R_{1}\leq 1+h(q)-h(r\star q)+h(r),\,R_{2}\leq 1+h(r),\nonumber\\
&&R_{1}\leq 2-h(p)-h(q),\,R_{2}\leq 2-h(p)-h(r\star q),\nonumber\\
&&R_{1}+R_{2}\leq 3-h(p)-h(q)-h(r)\},
\end{eqnarray}
\begin{eqnarray}\label{s9-r}
&&\mathcal{C}^{out}_{sf}=\{(R_{1}, R_{2}): R_{1}\leq 2-h(p)-h(q),\,R_{2}\leq 2-h(p)-h(r\star q),\nonumber\\
&&R_{1}+R_{2}\leq 3-h(p)-h(q)-h(r)\},
\end{eqnarray}
\begin{eqnarray}\label{s11-r}
&&\mathcal{C}^{in}_{s}=\{(R_{1}, R_{2}): R_{1}\leq 1-h(q),\,R_{2}\leq 1-h(r\star q)\},
\end{eqnarray}
where $a\star b=a(1-b)+(1-a)b$ and $h(a)=-a\log(a)-(1-a)\log(1-a)$ ($0\leq a\leq 1$).
Figure \ref{fr1} depicts $\mathcal{C}^{in-1}_{sf}$, $\mathcal{C}^{in-2}_{sf}$, $\mathcal{C}^{out}_{sf}$ and $\mathcal{C}^{in}_{s}$
for $p=q=r=0.05$. Form this figure, we conclude that the hybrid feedback strategy performs better than the secret key based feedback strategy, and both of
these strategies increase the secrecy rate region of the BC-MSR. Moreover, note that
there is still a gap between the inner and outer bounds on the secrecy capacity region of the BC-MSR with noiseless feedback.

\begin{figure}[htb]
\centering
\includegraphics[scale=0.4]{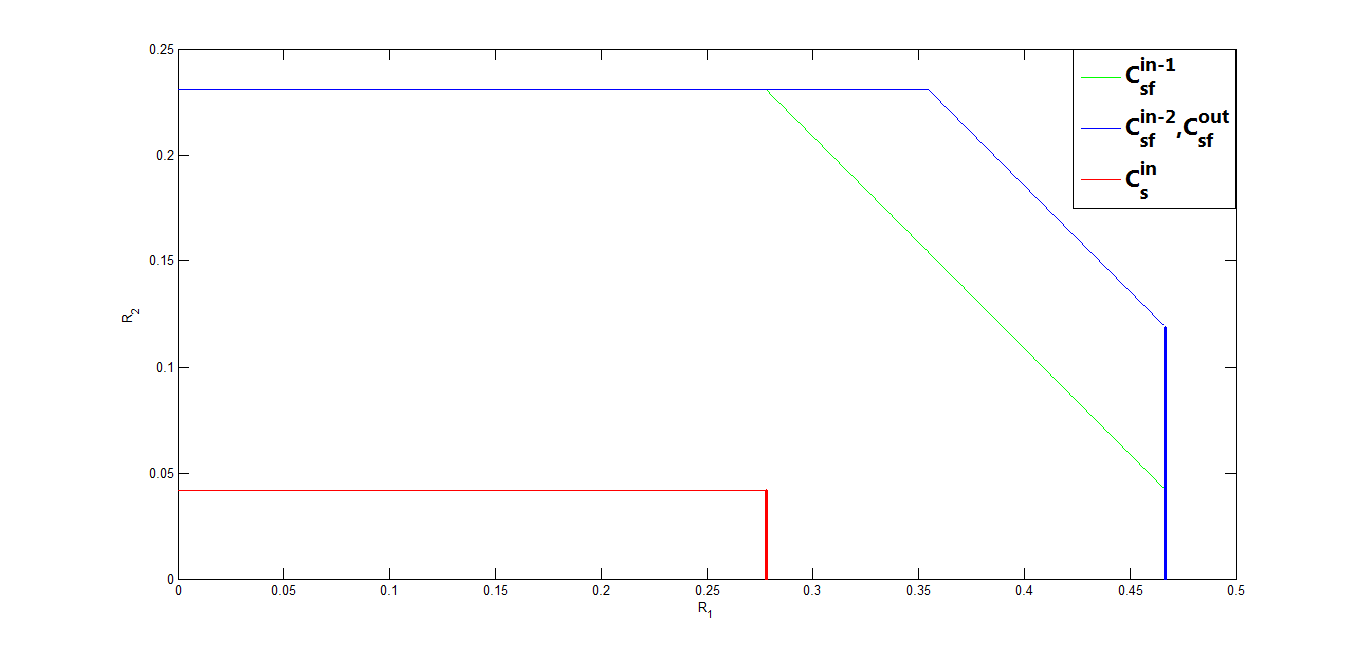}
\caption{Comparison of the bounds for the case $Z_{0}\rightarrow Z_{1}\rightarrow Z_{2}$ and $p=0.25$, $q=0.2$, $r=0.3$}
\label{fr2}
\end{figure}

Figure \ref{fr2} plots $\mathcal{C}^{in-1}_{sf}$, $\mathcal{C}^{in-2}_{sf}$, $\mathcal{C}^{out}_{sf}$ and $\mathcal{C}^{in}_{s}$
for $p=0.25$, $q=0.2$ and $r=0.3$. Form Figure \ref{fr2}, we conclude that for this case,
the secrecy capacity region of the BC-MSR with noiseless feedback is determined, and this is because the outer bound $\mathcal{C}^{out}_{sf}$ meets with the
hybrid strategy inner bound $\mathcal{C}^{in-2}_{sf}$.
Also, we see that the hybrid feedback coding strategy performs better than the secret key based feedback strategy, and both of them
increase the secrecy rate region of the BC-MSR.

\subsubsection{A special case of Dueck-type example with $Z_{1}\rightarrow Z_{0}\rightarrow Z_{2}$}\label{3-1-2}

\begin{figure}[htb]
\centering
\includegraphics[scale=0.4]{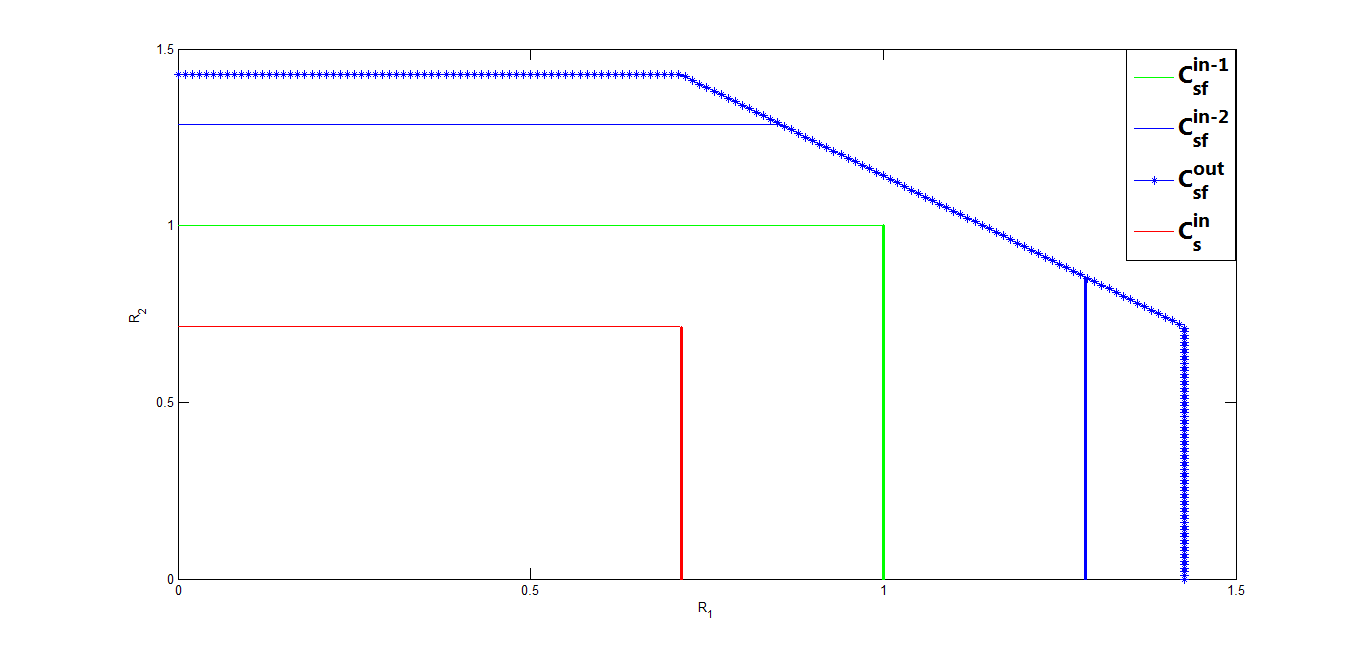}
\caption{Comparison of the bounds for the case $Z_{1}\rightarrow Z_{0}\rightarrow Z_{2}$ and $p=q=r=0.05$}
\label{fr3}
\end{figure}

For the case $Z_{1}\rightarrow Z_{0}\rightarrow Z_{2}$, define
\begin{eqnarray}\label{s12}
\left\{
\begin{array}{ll}
P(Z_{0}=0)=1-p,\,P(Z_{0}=1)=p,\\
P(Z_{1}=0|Z_{0}=0)=1-q,\,P(Z_{1}=1|Z_{0}=0)=q,\\
P(Z_{1}=0|Z_{0}=1)=q,\,P(Z_{1}=1|Z_{0}=1)=1-q,\\
P(Z_{2}=0|Z_{0}=0)=1-r,\,P(Z_{2}=1|Z_{0}=0)=r,\\
P(Z_{2}=0|Z_{0}=1)=r,\,P(Z_{2}=1|Z_{0}=1)=1-r
\end{array}
\right\},
\end{eqnarray}
where $0\leq p,q,r\leq \frac{1}{2}$. Substituting (\ref{s12}) and $Z_{0}\rightarrow Z_{1}\rightarrow Z_{2}$
into (\ref{s4}), (\ref{s5}), (\ref{s9}) and (\ref{s11}),
we have
\begin{eqnarray}\label{s4-r.1}
&&\mathcal{C}^{in-1}_{sf}=\{(R_{1}, R_{2}): R_{1}\leq 1,\,R_{2}\leq 1,\nonumber\\
&&R_{1}\leq 2-h(p)-h(q),\,R_{2}\leq 2-h(p)-h(r),\nonumber\\
&&R_{1}+R_{2}\leq 3-h(q)-h(p)-h(r)\},
\end{eqnarray}
\begin{eqnarray}\label{s5-r.1}
&&\mathcal{C}^{in-2}_{sf}=\{(R_{1}, R_{2}): R_{1}\leq 1+h(q),\,R_{2}\leq 1+h(r),\nonumber\\
&&R_{1}\leq 2-h(p)-h(q),\,R_{2}\leq 2-h(p)-h(r),\nonumber\\
&&R_{1}+R_{2}\leq 3-h(p)-h(q)-h(r)\},
\end{eqnarray}
\begin{eqnarray}\label{s9-r.1}
&&\mathcal{C}^{out}_{sf}=\{(R_{1}, R_{2}): R_{1}\leq 2-h(p)-h(q),\,R_{2}\leq 2-h(p)-h(r),\nonumber\\
&&R_{1}+R_{2}\leq 3-h(p)-h(q)-h(r)\},
\end{eqnarray}
\begin{eqnarray}\label{s11-r.1}
&&\mathcal{C}^{in}_{s}=\{(R_{1}, R_{2}): R_{1}\leq 1-h(q),\,R_{2}\leq 1-h(r)\}.
\end{eqnarray}
Figure \ref{fr3} depicts $\mathcal{C}^{in-1}_{sf}$, $\mathcal{C}^{in-2}_{sf}$, $\mathcal{C}^{out}_{sf}$ and $\mathcal{C}^{in}_{s}$
for $p=q=r=0.05$. It is easy to see that for the case $Z_{1}\rightarrow Z_{0}\rightarrow Z_{2}$,
the hybrid feedback strategy performs better than the secret key based feedback strategy, and both of
these strategies increase the secrecy rate region of the BC-MSR. Also 
notice that there is a gap between the inner and outer bounds.

\begin{figure}[htb]
\centering
\includegraphics[scale=0.4]{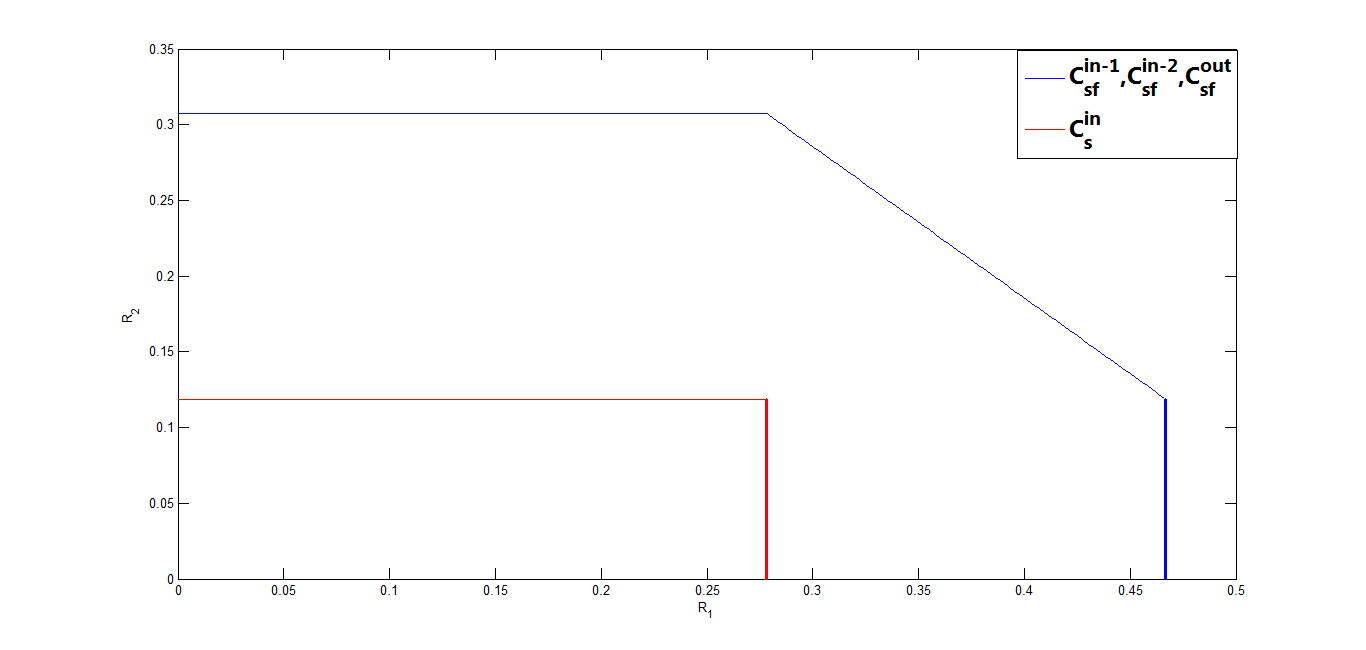}
\caption{Comparison of the bounds for the case $Z_{1}\rightarrow Z_{0}\rightarrow Z_{2}$ and $p=0.25$, $q=0.2$, $r=0.3$}
\label{fr4}
\end{figure}

Figure \ref{fr4} plots $\mathcal{C}^{in-1}_{sf}$, $\mathcal{C}^{in-2}_{sf}$, $\mathcal{C}^{out}_{sf}$ and $\mathcal{C}^{in}_{s}$
for $p=0.25$, $q=0.2$ and $r=0.3$. Form Figure \ref{fr4}, we conclude that for this case,
the secrecy capacity region of the BC-MSR with noiseless feedback is determined, and it equals to the inner bounds
$\mathcal{C}^{in-1}_{sf}$, $\mathcal{C}^{in-2}_{sf}$, and the outer bound $\mathcal{C}^{out}_{sf}$.
Also, we see that for this case, the secret key based feedback strategy performs as well as the hybrid feedback coding strategy,
and both of them
increase the secrecy rate region of the BC-MSR.

\subsection{Blackwell-type Example}\label{3-2}

\begin{figure}[htb]
\centering
\includegraphics[scale=0.4]{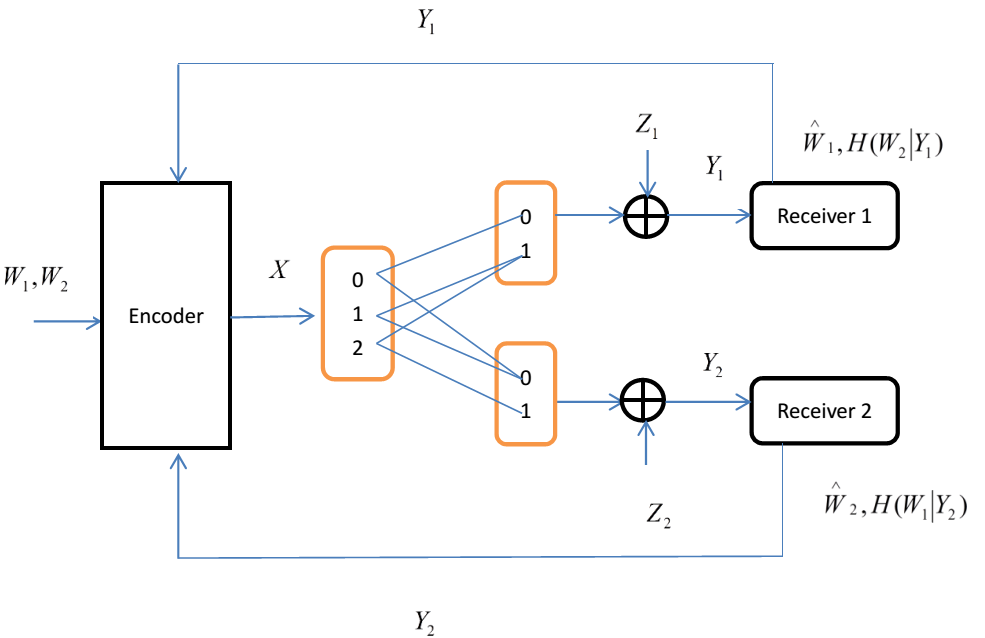}
\caption{A Blackwell-type example of the BC-MSR with noiseless feedback}
\label{f2-dota}
\end{figure}

In this subsection, we explain the inner and outer bounds on the secrecy capacity region of the BC-MSR with noiseless feedback via a Blackwell-type example.
In this example (see Figure \ref{f2-dota}), the channel input $X$ chooses values from $\{0,1,2\}$, the channel outputs $Y_{1}$ and $Y_{2}$ choose values from
$\{0,1\}$,
and they satisfy
\begin{eqnarray}\label{s3-d1}
\left\{
\begin{array}{ll}
X=0,\,\,\,Y_{1}=Z_{1},\,\,\,Y_{2}=Z_{2},\\
X=1,\,\,\,Y_{1}=1\oplus Z_{1},\,\,\,Y_{2}=Z_{2},\\
X=2,\,\,\,Y_{1}=1\oplus Z_{1},\,\,\,Y_{2}=1\oplus Z_{2},
\end{array}
\right\},
\end{eqnarray}
where $\oplus$ is the modulo addition over $\{0,1\}$, the noises $Z_{1}\sim Bern(p)$, $Z_{2}\sim Bern(p)$ ($0\leq p\leq 0.5$),
and the noises $Z_{1}$, $Z_{2}$ are mutually independent and they are independent of the channel inputs.

For this Blackwell-type example, the secret key based inner bound $\mathcal{C}^{f-in-1*}_{s}$ is obtained by defining $P(Q=0)=P(Q=1)=\frac{1}{2}$,
\begin{eqnarray}\label{dac1}
&&P(U_{1}=0,U_{2}=0|Q=0)=\alpha,\,P(U_{1}=0,U_{2}=1|Q=0)=0,\nonumber\\
&&P(U_{1}=1,U_{2}=0|Q=0)=1-\alpha-\beta,\,P(U_{1}=1,U_{2}=1|Q=0)=\beta,\nonumber\\
&&P(U_{1}=0,U_{2}=0|Q=1)=\beta,\,P(U_{1}=0,U_{2}=1|Q=1)=0,\nonumber\\
&&P(U_{1}=1,U_{2}=0|Q=1)=1-\alpha-\beta,\,P(U_{1}=1,U_{2}=1|Q=1)=\alpha,\nonumber\\
\end{eqnarray}
$X=U_{1}+U_{2}$, and substituting (\ref{s3-d1}), $Z_{1}\sim Bern(p)$ and $Z_{2}\sim Bern(p)$ into Theorem \ref{T1}. Hence we have
\begin{eqnarray}\label{dac3}
&&\mathcal{C}^{f-in-1*}_{s}=\{(R_{1}, R_{2}): \nonumber\\
&&R_{1}\leq \frac{1}{2}h(\alpha\star p)+\frac{1}{2}h(\beta\star p)
-(1-\beta)\log\frac{1}{1-\beta}-(1-\alpha)\log\frac{1}{1-\alpha}+(1-\alpha-\beta)\log\frac{1}{1-\alpha-\beta},\nonumber\\
&&R_{2}\leq \frac{1}{2}h(\alpha\star p)+\frac{1}{2}h(\beta\star p)
-(1-\beta)\log\frac{1}{1-\beta}-(1-\alpha)\log\frac{1}{1-\alpha}+(1-\alpha-\beta)\log\frac{1}{1-\alpha-\beta},\nonumber\\
&&R_{1}\leq \frac{1}{2}h(\alpha\star p)+\frac{1}{2}h(\beta\star p)-h(p),\nonumber\\
&&R_{2}\leq \frac{1}{2}h(\alpha\star p)+\frac{1}{2}h(\beta\star p)-h(p),\nonumber\\
&&R_{1}+R_{2}\leq h(\alpha\star p)+h(\beta\star p)-(1-\beta)\log\frac{1}{1-\beta}-(1-\alpha)\log\frac{1}{1-\alpha}+(1-\alpha-\beta)\log\frac{1}{1-\alpha-\beta}\},\nonumber\\
\end{eqnarray}
where $a\star b=a(1-b)+(1-a)b$ and $h(a)=-a\log(a)-(1-a)\log(1-a)$ ($0\leq a\leq 1$).

Then, using the above definitions for $\mathcal{C}^{f-in-1*}_{s}$ and letting $V_{0}=Z$, $V_{1}=U_{1}$, $V_{2}=U_{2}$,
the hybrid inner bound $\mathcal{C}^{f-in-2*}_{s}$ is obtained by substituting (\ref{s3-d1}), $Z_{1}\sim Bern(p)$ and
$Z_{2}\sim Bern(p)$ into Theorem \ref{T2}, and it is given by
\begin{eqnarray}\label{dac4}
&&\mathcal{C}^{f-in-2*}_{s}=\{(R_{1}, R_{2}): \nonumber\\
&&R_{1}\leq \min\{\frac{1}{2}h(\alpha)+\frac{1}{2}h(\beta), h(\alpha,\beta,1-\alpha-\beta)-\frac{1}{2}h(\alpha)-\frac{1}{2}h(\beta)+h(p)\},\nonumber\\
&&R_{2}\leq \min\{\frac{1}{2}h(\alpha)+\frac{1}{2}h(\beta), h(\alpha,\beta,1-\alpha-\beta)-\frac{1}{2}h(\alpha)-\frac{1}{2}h(\beta)+h(p)\},\nonumber\\
&&R_{1}\leq h(\frac{\alpha+\beta}{2}\star p)-2h(p),\,\,R_{2}\leq h(\frac{\alpha+\beta}{2}\star (1-p))-2h(p)\nonumber\\
&&R_{1}+R_{2}\leq h(\frac{\alpha+\beta}{2}\star p)-2h(p)-\frac{1}{2}h(\alpha)-\frac{1}{2}h(\beta)+h(\alpha,\beta,1-\alpha-\beta)\},\nonumber\\
\end{eqnarray}
where $h(\alpha,\beta,1-\alpha-\beta)=-\alpha\log(\alpha)-\beta\log(\beta)-(1-\alpha-\beta)\log(1-\alpha-\beta)$.

Next, we show an outer bound $\mathcal{C}^{f-out*}_{s}$ on the secrecy capacity region of this Blackwell-type example, and it is obtained from
Theorem \ref{T3}. Specifically, the bound
$R_{1}\leq \min\{I(U_{1};Y_{1}|Q)-I(U_{1};Y_{2}|Q), I(U_{1};Y_{1}|Q,U_{2})-I(U_{1};Y_{2}|Q,U_{2}),H(Y_{1}|Q,U_{2},Y_{2})\}$ in Theorem \ref{T3} can be further bounded by
\begin{eqnarray}\label{dac5}
&&I(U_{1};Y_{1}|Q)-I(U_{1};Y_{2}|Q)\nonumber\\
&&\leq I(U_{1};Y_{1}|Q)\leq H(Y_{1})-H(Y_{1}|Q,U_{1})\nonumber\\
&&\leq H(Y_{1})-H(Y_{1}|Q,U_{1},X)\stackrel{(1)}=H(Y_{1})-H(Y_{1}|X)=I(X;Y_{1}),
\end{eqnarray}
\begin{eqnarray}\label{dac6}
&&I(U_{1};Y_{1}|Q,U_{2})-I(U_{1};Y_{2}|Q,U_{2})\nonumber\\
&&\leq I(U_{1};Y_{1}|Q,U_{2})\leq H(Y_{1})-H(Y_{1}|Q,U_{1},U_{2})\nonumber\\
&&\leq H(Y_{1})-H(Y_{1}|Q,U_{1},U_{2},X)\stackrel{(2)}=H(Y_{1})-H(Y_{1}|X)=I(X;Y_{1}),
\end{eqnarray}
\begin{eqnarray}\label{dac7}
&&H(Y_{1}|Q,U_{2},Y_{2})\leq H(Y_{1}|Y_{2}),
\end{eqnarray}
where (1) is from $(Q,U_{1})\rightarrow X\rightarrow Y_{1}$, and (2) is from $(Q,U_{1},U_{2})\rightarrow X\rightarrow Y_{1}$.
Hence we have $R_{1}\leq \min\{I(X;Y_{1}),H(Y_{1}|Y_{2})\}$.
Analogously, we have $R_{2}\leq \min\{I(X;Y_{2}),H(Y_{2}|Y_{1})\}$.
Now defining
\begin{eqnarray}\label{dac8}
&&P(X=0)=\alpha_{1},\,\,P(X=1)=\alpha_{2},\,\,P(X=2)=1-\alpha_{1}-\alpha_{2},
\end{eqnarray}
and substituting (\ref{s3-d1}), $Z_{1}\sim Bern(p)$ and
$Z_{2}\sim Bern(p)$ into $R_{1}\leq \min\{I(X;Y_{1}),H(Y_{1}|Y_{2})\}$ and $R_{2}\leq \min\{I(X;Y_{2}),H(Y_{2}|Y_{1})\}$,
the outer bound $\mathcal{C}^{f-out*}_{s}$ is given by
\begin{eqnarray}\label{dac9}
&&\mathcal{C}^{f-out*}_{s}=\{(R_{1}, R_{2}): \nonumber\\
&&R_{1}\leq \min\{h(\alpha_{1}\star p)-h(p),A-h(\alpha_{1}+\alpha_{2})\star p)\},\nonumber\\
&&R_{2}\leq \min\{h((\alpha_{1}+\alpha_{2})\star p)-h(p),A-h(\alpha_{1}\star p)\},
\end{eqnarray}
where
\begin{eqnarray}\label{dac10}
&&A=-(\alpha_{1}\bar{p}^{2}+\alpha_{2}p\bar{p}+\overline{\alpha_{1}+\alpha_{2}}p^{2})\log(\alpha_{1}\bar{p}^{2}+\alpha_{2}p\bar{p}+\overline{\alpha_{1}+\alpha_{2}}p^{2})\nonumber\\
&&-(\alpha_{1}\bar{p}p+\alpha_{2}p^{2}+\overline{\alpha_{1}+\alpha_{2}}p\bar{p})\log(\alpha_{1}\bar{p}p+\alpha_{2}p^{2}+\overline{\alpha_{1}+\alpha_{2}}p\bar{p})\nonumber\\
&&-(\alpha_{1}\bar{p}p+\alpha_{2}\bar{p}^{2}+\overline{\alpha_{1}+\alpha_{2}}p\bar{p})\log(\alpha_{1}\bar{p}p+\alpha_{2}\bar{p}^{2}+\overline{\alpha_{1}+\alpha_{2}}p\bar{p})\nonumber\\
&&-(\alpha_{1}p^{2}+\alpha_{2}p\bar{p}+\overline{\alpha_{1}+\alpha_{2}}\bar{p}^{2})\log(\alpha_{1}p^{2}+\alpha_{2}p\bar{p}+\overline{\alpha_{1}+\alpha_{2}}\bar{p}^{2}),
\end{eqnarray}
$\bar{p}=1-p$ and $\overline{\alpha_{1}+\alpha_{2}}=1-\alpha_{1}-\alpha_{2}$.

Finally, we show the inner bound $\mathcal{C}^{in**}_{s}$ on the secrecy capacity region
of this Blackwell-type example without the feedback, and it is obtained by using the above definitions for $\mathcal{C}^{f-in-1*}_{s}$ and substituting
(\ref{s3-d1}), $Z_{1}\sim Bern(p)$ and $Z_{2}\sim Bern(p)$ into (\ref{s10}). Hence we have
\begin{eqnarray}\label{dac11}
&&\mathcal{C}^{in**}_{s}=\{(R_{1}, R_{2}): \nonumber\\
&&R_{1}\leq \frac{1}{2}h(\alpha\star p)+\frac{1}{2}h(\beta\star p)-h(p)
-\bar{\beta}\log\frac{1}{\bar{\beta}}-\bar{\alpha}\log\frac{1}{\bar{\alpha}}-\overline{\alpha+\beta}\log\overline{\alpha+\beta},\nonumber\\
&&R_{2}\leq \frac{1}{2}h(\alpha\star p)+\frac{1}{2}h(\beta\star p)-h(p)
-\bar{\beta}\log\frac{1}{\bar{\beta}}-\bar{\alpha}\log\frac{1}{\bar{\alpha}}-\overline{\alpha+\beta}\log\overline{\alpha+\beta}\},
\end{eqnarray}
where $\bar{\beta}=1-\beta$, $\bar{\alpha}=1-\alpha$ and $\overline{\alpha+\beta}=1-\alpha-\beta$.

\begin{figure}[htb]
\centering
\includegraphics[scale=0.4]{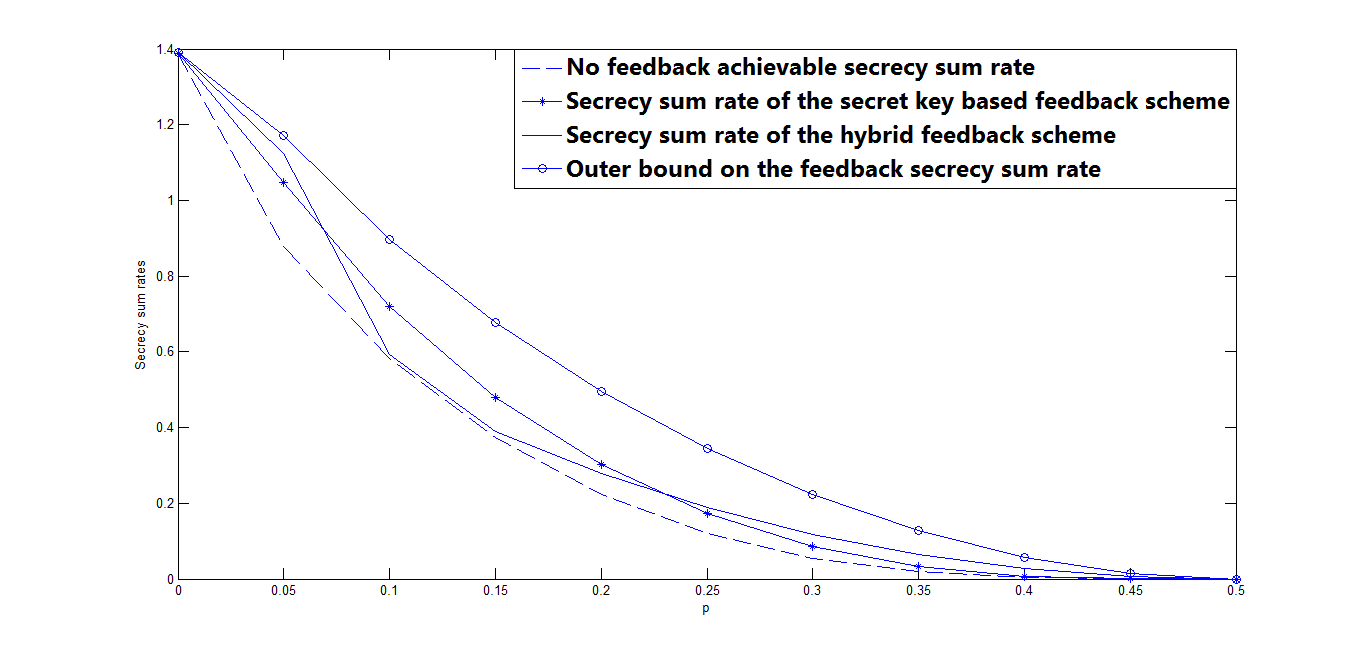}
\caption{The maximum secrecy sum rates of the bounds for the Blackwell-type example of the BC-MSR with noiseless feedback}
\label{fm-1}
\end{figure}

To show which feedback strategy performs better in enhancing the total secrecy rate of the BC-MSR, in this Blackwell-type example, we depict the secrecy sum rates
of the bounds $\mathcal{C}^{f-in-1*}_{s}$, $\mathcal{C}^{f-in-2*}_{s}$, $\mathcal{C}^{f-out*}_{s}$
and $\mathcal{C}^{in**}_{s}$ with different values of $p$ (here $p$ is w.r.t. the channel noise), see Figure \ref{fm-1}.
From Figure \ref{fm-1}, we see that both feedback strategies increase the secrecy sum rate of $\mathcal{C}^{in**}_{s}$,
and the hybrid feedback strategy
does not always perform better than the secret key based feedback strategy in enhancing the secrecy sum rate of this Blackwell-type BC-MSR.
Moreover, as shown in Figure \ref{fm-1}, there exists a gap between the secrecy sum rates
of the inner and outer bounds, and eliminating this gap is still a tough work.

\section{Conclusion\label{secVI}}

Two feedback coding strategies for the BC-MSR are proposed, and the comparison of the two strategies is shown via two examples.
Specifically, in the Dueck-type example, we show the achievable secrecy rate region of the secret key based feedback strategy
is no larger than that of the hybrid feedback strategy. For the Blackwell-type example, we show the secrecy sum rate of the
hybrid feedback strategy is not always larger than that of the secret key based feedback strategy.
From these two examples, we see that for the BC-MSR,
our new hybrid feedback strategy may perform better than the already existing secret key based feedback strategy, which offers a new option for
enhancing the physical layer security of the broadcast channel models.

\renewcommand{\theequation}{A\arabic{equation}}
\appendices\section{Proof of Theorem \ref{T3}\label{rotk1}}
\setcounter{equation}{0}

First, define $Y_{1}^{i-1}=(Y_{1,1},Y_{1,2},...,Y_{1,i-1})$, $Y_{1,i+1}^{N}=(Y_{1,i+1},Y_{1,i+2},...,Y_{1,N})$,
$Y_{2}^{i-1}=(Y_{2,1},Y_{2,2},...,Y_{2,i-1})$ and $Y_{2,i+1}^{N}=(Y_{2,i+1},Y_{2,i+2},...,Y_{2,N})$. Further define
\begin{eqnarray}\label{b1}
&&Q\triangleq (Y_{1}^{L-1}, Y_{2,L+1}^{N}, L), U_{1}\triangleq (Q, W_{1}), U_{2}\triangleq (Q, W_{2}), Y_{1}\triangleq Y_{1,L}, Y_{2}\triangleq Y_{2,L},
\end{eqnarray}
where the RV $L$ is used for time sharing, and it is randomly drawn from the set $\{1,2,...,N\}$. In addition, $L$ is independent of the channel input and outputs. 

Next, using the above definitions, we show that $R_{1}$ can be upper bounded by 
\begin{eqnarray}\label{b2}
&&R_{1}-\epsilon\stackrel{(1)}\leq\frac{1}{N}H(W_{1}|Y_{2}^{N})\nonumber\\
&&=\frac{1}{N}(H(W_{1}|Y_{2}^{N},W_{2})+I(W_{1};W_{2}|Y_{2}^{N}))\nonumber\\
&&\stackrel{(2)}\leq \frac{1}{N}H(W_{1}|Y_{2}^{N},W_{2})+\frac{\delta(\epsilon)}{N}\nonumber\\
&&=\frac{1}{N}(H(W_{1}|W_{2})-I(W_{1};Y_{2}^{N}|W_{2}))+\frac{\delta(\epsilon)}{N}\nonumber\\
&&\stackrel{(3)}\leq\frac{1}{N}(I(W_{1};Y_{1}^{N}|W_{2})-I(W_{1};Y_{2}^{N}|W_{2}))+\frac{2\delta(\epsilon)}{N}\nonumber\\
&&=\frac{1}{N}\sum_{i=1}^{N}(I(W_{1};Y_{1,i}|Y_{1}^{i-1},W_{2})-I(W_{1};Y_{2,i}|Y_{2,i+1}^{N},W_{2}))+\frac{2\delta(\epsilon)}{N}\nonumber\\
&&=\frac{1}{N}\sum_{i=1}^{N}(I(W_{1};Y_{1,i}|Y_{1}^{i-1},W_{2},Y_{2,i+1}^{N})+I(Y_{1,i};Y_{2,i+1}^{N}|W_{2},Y_{1}^{i-1})
-I(Y_{1,i};Y_{2,i+1}^{N}|W_{1},W_{2},Y_{1}^{i-1})\nonumber\\
&&-I(W_{1};Y_{2,i}|Y_{1}^{i-1},W_{2},Y_{2,i+1}^{N})-I(Y_{2,i};Y_{1}^{i-1}|W_{2},Y_{2,i+1}^{N})+I(Y_{2,i};Y_{1}^{i-1}|W_{1},W_{2},Y_{2,i+1}^{N}))
+\frac{2\delta(\epsilon)}{N}\nonumber\\
&&\stackrel{(4)}=\frac{1}{N}\sum_{i=1}^{N}(I(W_{1};Y_{1,i}|Y_{1}^{i-1},W_{2},Y_{2,i+1}^{N})-I(W_{1};Y_{2,i}|Y_{1}^{i-1},W_{2},Y_{2,i+1}^{N}))
+\frac{2\delta(\epsilon)}{N}\nonumber\\
&&\stackrel{(5)}=I(W_{1};Y_{1,L}|Y_{1}^{L-1},W_{2},Y_{2,L+1}^{N},L)-I(W_{1};Y_{2,L}|Y_{1}^{L-1},W_{2},Y_{2,L+1}^{N},L)+\frac{2\delta(\epsilon)}{N}\nonumber\\
&&\stackrel{(6)}=I(U_{1};Y_{1}|U_{2},Q)-I(U_{1};Y_{2}|U_{2},Q)+\frac{2\delta(\epsilon)}{N},
\end{eqnarray}
where (1) follows by (\ref{a3}),
(2) and (3) follow by Fano's inequality, $P_{e,1}\leq \epsilon$ and $P_{e,2}\leq \epsilon$, (4) is from
the Csisz$\acute{a}$r's equalities
\begin{eqnarray}\label{b1-ma1}
&&I(Y_{1,i};Y_{2,i+1}^{N}|W_{2},Y_{1}^{i-1})=I(Y_{2,i};Y_{1}^{i-1}|W_{2},Y_{2,i+1}^{N}),
\end{eqnarray}
and
\begin{eqnarray}\label{b1-ma2}
&&I(Y_{1,i};Y_{2,i+1}^{N}|W_{1},W_{2},Y_{1}^{i-1})=I(Y_{2,i};Y_{1}^{i-1}|W_{1},W_{2},Y_{2,i+1}^{N}),
\end{eqnarray}
(5) follows by the definition of $J$,
and (6) follows by (\ref{b1}). Letting $\epsilon\rightarrow 0$, we get $R_{1}\leq I(U_{1};Y_{1}|U_{2},Q)-I(U_{1};Y_{2}|U_{2},Q)$.

Moreover, note that 
\begin{eqnarray}\label{b3}
&&R_{1}-\epsilon\leq\frac{1}{N}H(W_{1}|Y_{2}^{N})\nonumber\\
&&=\frac{1}{N}(H(W_{1})-I(W_{1};Y_{2}^{N}))\nonumber\\
&&\stackrel{(a)}\leq \frac{1}{N}(I(W_{1};Y_{1}^{N})-I(W_{1};Y_{2}^{N}))+\frac{\delta(\epsilon)}{N}\nonumber\\
&&=\frac{1}{N}\sum_{i=1}^{N}(I(W_{1};Y_{1,i}|Y_{1}^{i-1})-I(W_{1};Y_{2,i}|Y_{2,i+1}^{N}))+\frac{\delta(\epsilon)}{N}\nonumber\\
&&=\frac{1}{N}\sum_{i=1}^{N}(I(W_{1};Y_{1,i}|Y_{1}^{i-1},Y_{2,i+1}^{N})+I(Y_{1,i};Y_{2,i+1}^{N}|Y_{1}^{i-1})
-I(Y_{1,i};Y_{2,i+1}^{N}|W_{1},Y_{1}^{i-1})\nonumber\\
&&-I(W_{1};Y_{2,i}|Y_{1}^{i-1},Y_{2,i+1}^{N})-I(Y_{2,i};Y_{1}^{i-1}|Y_{2,i+1}^{N})+I(Y_{2,i};Y_{1}^{i-1}|W_{1},Y_{2,i+1}^{N}))
+\frac{\delta(\epsilon)}{N}\nonumber\\
&&\stackrel{(b)}=\frac{1}{N}\sum_{i=1}^{N}(I(W_{1};Y_{1,i}|Y_{1}^{i-1},Y_{2,i+1}^{N})-I(W_{1};Y_{2,i}|Y_{1}^{i-1},Y_{2,i+1}^{N}))
+\frac{\delta(\epsilon)}{N}\nonumber\\
&&\stackrel{(c)}=I(W_{1};Y_{1,L}|Y_{1}^{L-1},Y_{2,L+1}^{N},L)-I(W_{1};Y_{2,L}|Y_{1}^{L-1},Y_{2,L+1}^{N},L)+\frac{\delta(\epsilon)}{N}\nonumber\\
&&\stackrel{(d)}=I(U_{1};Y_{1}|Q)-I(U_{1};Y_{2}|Q)+\frac{\delta(\epsilon)}{N},
\end{eqnarray}
where (a) follows by Fano's inequality and $P_{e,1}\leq \epsilon$, (b) follows from the similar Csisz$\acute{a}$r's equalities of (\ref{b1-ma1})
and (\ref{b1-ma2}), (c) follows by the definition of $J$,
and (d) follows by (\ref{b1}). Letting $\epsilon\rightarrow 0$, we get $R_{1}\leq I(U_{1};Y_{1}|Q)-I(U_{1};Y_{2}|Q)$.

Finally, notice that
\begin{eqnarray}\label{b4}
&&R_{1}-\epsilon\leq\frac{1}{N}H(W_{1}|Y_{2}^{N})\nonumber\\
&&=\frac{1}{N}(H(W_{1}|Y_{2}^{N},W_{2})+I(W_{1};W_{2}|Y_{2}^{N}))\nonumber\\
&&\leq \frac{1}{N}H(W_{1}|Y_{2}^{N},W_{2})+\frac{\delta(\epsilon)}{N}\nonumber\\
&&=\frac{1}{N}(H(W_{1}|Y_{2}^{N},W_{2})-H(W_{1}|Y_{2}^{N},W_{2},Y_{1}^{N})+H(W_{1}|Y_{2}^{N},,W_{2},Y_{1}^{N}))+\frac{\delta(\epsilon)}{N}\nonumber\\
&&\stackrel{(e)}\leq \frac{1}{N}I(W_{1};Y_{1}^{N}|Y_{2}^{N},W_{2})+\frac{2\delta(\epsilon)}{N}\nonumber\\
&&\leq \frac{1}{N}H(Y_{1}^{N}|Y_{2}^{N},W_{2})+\frac{2\delta(\epsilon)}{N}\nonumber\\
&&=\frac{1}{N}\sum_{i=1}^{N}H(Y_{1,i}|Y_{1}^{i-1},Y_{2}^{N},W_{2})+\frac{2\delta(\epsilon)}{N}\nonumber\\
&&\leq \frac{1}{N}\sum_{i=1}^{N}H(Y_{1,i}|Y_{1}^{i-1},Y_{2,i+1}^{N},Y_{2,i},W_{2})+\frac{2\delta(\epsilon)}{N}\nonumber\\
&&\stackrel{(f)}=H(Y_{1,L}|Y_{1}^{L-1},Y_{2,L+1}^{N},Y_{2,L},W_{2},L)+\frac{2\delta(\epsilon)}{N}\nonumber\\
&&\stackrel{(g)}=H(Y_{1}|Q,U_{2},Y_{2})+\frac{2\delta(\epsilon)}{N},
\end{eqnarray}
where (e) follows by Fano's inequality,
(f) follows by the definition of $J$,
and (g) follows by (\ref{b1}).
Letting $\epsilon\rightarrow 0$, we get $R_{1}\leq H(Y_{1}|Q,U_{2},Y_{2})$.
Now the proof of $R_{1}\leq \min\{I(U_{1};Y_{1}|Q)-I(U_{1};Y_{2}|Q), I(U_{1};Y_{1}|Q,U_{2})-I(U_{1};Y_{2}|Q,U_{2}),H(Y_{1}|Q,U_{2},Y_{2})\}$ is completed.
Analogously, we can prove $R_{2}\leq \min\{I(U_{2};Y_{2}|Q)-I(U_{2};Y_{1}|Q), I(U_{2};Y_{2}|Q,U_{1})-I(U_{2};Y_{1}|Q,U_{1}),
H(Y_{2}|Q,U_{1},Y_{1})\}$.
The proof of Theorem \ref{T3} is completed.

\end{document}